\documentclass{aa}  

\usepackage{graphicx}
\usepackage{txfonts}

\usepackage{amsmath}
\usepackage{amssymb}
\usepackage{graphicx}

\usepackage{color}

\usepackage{natbib}
\newcommand{\Mpch}{\,{\rm Mpc}\,\ifmmode h^{-1}\else $h^{-1}$\fi}
\newcommand{\Msh}{\,\ifmmode M_\odot\,h^{-1}\else $M_\odot\,h^{-1}$\fi}
\newcommand{\kms}{\,\mathrm{km}\,\mathrm{s}^{-1}}

\newcommand{\fofrfour}{FofR04\ }
\newcommand{\fofrfive}{FofR05\ }
\newcommand{\fofrsix}{FofR06\ }
\newcommand{\symmA}{Symmetron A\ }
\newcommand{\symmB}{Symmetron B\ }
\newcommand{\symmC}{Symmetron C\ }
\newcommand{\symmD}{Symmetron D\ }

\newcommand{\BothPointOne}{DM0.1G0.1}
\newcommand{\BothOne}{DM1G1}

\newcommand{\DMPointOne}{DM0.1G1}

\newcommand{\GasPointOne}{DM1G0.1}

\newcommand{\fig}[1]{Fig.~\ref{#1}}

\begin{document}

\title{Estimates of cluster masses in screened modified gravity}

\author{M. Gronke
          \inst{1}
          \and
          A. Hammami
          \inst{1}
          \and
          D. F. Mota\inst{1}
          \and
          H. A. Winther\inst{2}
}

   \institute{Institute of Theoretical Astrophysics, University of Oslo, Postboks 1029, 0315 Oslo, Norway\\
              \email{maxbg@astro.uio.no}
         \and
            Astrophysics, University of Oxford, DWB, Keble Road, Oxford, OX1 3RH, UK\\
             }

   \date{Received 27 May 2016 / Accepted 9 September 2016}

  \abstract
{
We use cosmological hydrodynamical simulations to study the effect of screened modified gravity models on the mass estimates of galaxy clusters. In particular, we focus on two novel aspects: \textit{(i)} we study modified gravity models in which baryons and dark matter are coupled with different strengths to the scalar field, and, \textit{(ii)} we put the simulation results into the greater context of a general screened-modified gravity parametrization. We have compared the mass of clusters inferred via lensing versus the mass inferred via kinematical measurements as a probe of violations of the equivalence principle at Mpc scales. We find that estimates of cluster masses via X-ray observations is mainly sensitive to the coupling between the scalar degree of freedom and baryons -- while the kinematical mass is mainly sensitive to the coupling to dark matter. Therefore, the relation between the two mass estimates is a probe of a possible non-universal coupling between the scalar field, the standard model fields, and dark matter. Finally, we used observational data of kinetic, thermal and lensing masses to place constraints on deviations from general relativity on cluster scales for a general parametrization of screened modified gravity theories which contains $f(R)$ and Symmetron models.
We find that while the kinematic mass can be used to place competitive constraints, using thermal measurements is challenging as a potential non-thermal contribution is degenerate with the imprint of modified gravity.
}


   \maketitle

\section{Introduction}
Over a decade has passed since the indisputable discovery of the accelerated expansion of the Universe \citep{Riess1998AJ....116.1009R,Perlmutter1999ApJ...517..565P} but its physical origin is still unknown. A possible -- and rather popular -- solution is to modify the theory of general relativity (GR). This has been done for a number of years and lead to numerous theories of modified gravity \citep[for reviews, see, e.g.,][]{Amendola2010deto.book.....A,Clifton2012}.
The main challenge for many modified gravity theories are measurements of the gravitational strength on Earth and in the solar system \citep[e.g.,][]{Berotti2003Natur.425..374B, Will2006LRR.....9....3W, Williams2004PhRvL..93z1101W}, which confirm the predictions of GR with great precision. One viable solution to this is to employ a so-called screening mechanism which restores GR in the solar system. Screening mechanisms are usually triggered by large local matter density or space-time curvature and lead to a convergence of the gravitational strength to its value predicted by GR.

For the sub-category of the extension of GR in the scalar sector\footnote{Also, other extensions of GR, for example, in the vectorial sector are possible. However, apart from managing theoretical difficulties they are also obliged not to violate the local constraints mentioned, and, thus might also employ a screening mechanism.}, that is, by adding a coupled scalar field to the Einstein-Hilbert action, several possible screening mechanisms are on the market \citep[see, e.g.,][]{Khoury2010,Joyce2014}. They can be categorized as follows:
\begin{itemize}
\item \textit{Screening because of the scalar field value} -- also often denoted as Chameleon screening. This group can be further divided into screening mechanisms that affect directly the coupling strength -- such as the Dilaton \citep{Damour1994NuPhB.423..532D} and the Symmetron \citep{Hinterbichler,Hinterbichlera} screening -- as well as mechanisms that alter the range of the additional force. The latter screening is often dubbed Chameleon screening \citep{Khourya,2004PhRvD..69d4026K,gan,2007PhRvD..75f3501M}.
\item \textit{Screening due to derivatives of the field value} -- also called Vainshtein-like screening. Here, one can differentiate between screening due to the first or the second derivative of the scalar field. Screening mechanisms belonging to the former group are the k-Mouflage \citep{2009IJMPD..18.2147B,mota3,2014PhRvD..90b3507B} and D-Bionic screening \citep{BurragePhysRevD.90.024001} whereas the latter group consists of the eponymous Vainshtein screening \citep{1972PhLB...39..393V}.
\end{itemize}
  It is important to differentiate between the screening mechanism and the particular theory of gravity employing this mechanism. For instance, particular theories employing the Vainshtein screening are the DGP model \citep{2000PhLB..485..208D}, Galileons \citep{2009PhRvD..79f4036N}, and, massive gravity \citep{2014LRR....17....7D}.

This wealth of theoretical alternatives to GR stands in stark contrast to the observational findings which, so far, confirm GR on a variety of environments \& scales \citep[for observational reviews see, e.g.,][]{2015arXiv150404623K,Baker2015ApJ...802...63B,2015arXiv151205356B} although deviations in many observables are predicted. Apart from the background cosmology \citep[e.g.,][for the Vainshtein, Chameleon and Symmetron, respectively]{2014PhRvD..90l4014K,2004PhRvD..70l3518B,Hinterbichlera} usually $N$-body codes are used to study screened modified gravity models \citep[for a review of the numerical techniques, see][]{Winther2015arXiv150606384W}. The most common approach is to start a $\Lambda$CDM and a modified gravity simulation using the same initial conditions and then analyze the deviations between the simulation outputs at later times. In this way, deviations in the matter power spectrum  \citep{2008PhRvD..78l3524O,mota4,Li2012JCAP...01..051L,bour,2013JCAP...05..023L,2013PhRvL.110p1101L,Puchwein2013MNRAS.436..348P}, the halo mass function \citep{2010PhRvD..81j3002S,2013JCAP...10..027B,mota2,Davis2012ApJ...748...61D,2015arXiv151101494A}, the velocity field \citep{2014arXiv1408.2856C,2014PhRvL.112v1102H,Gronke2014b_dl,2016arXiv160303072F}, gravitational lensing \citep{2015MNRAS.454.4085B,2015JCAP...10..036T,2016arXiv160301325H} and many other quantities have been found. These predictions give valuable insights into the way in which mechanisms act on the environment. However, exactly how transferable to observations they are, is questionable due to the neglecting of baryonic effects which are somewhat degenerate with the enhancement of gravity \citep{Puchwein2013MNRAS.436..348P,2014MNRAS.440..833A,Hammami2015a}, and, more importantly the direct comparison with another, alternative `Universe' -- a technique which is certainly not possible with real data.

Another problem associated with the confrontation of the numerical predictions with real data is the richness of the modified gravity landscape. Not only is the above mentioned number of models incomplete (and steadily increasing) but each model has its own (often multi-dimensional) parameter space. This makes the classical approach, by which we mean, using a suite of $N$-body simulations to constrain the model parameter spaces one-by-one, unfeasible. One alternative is to speed up the numerical simulations tremendously as done by \citet{Mead2014}, \citet{2015JCAP...12..059B}  and \citet{2014arXiv1403.6492W}. Alternatively, one can try to unify the predictions of several modified gravity models potentially allowing to rule out (parts of parameter spaces) of several models at once. This path was taken theoretically by \citet{2012PhLB..715...38B,2012PhRvD..86d4015B} who developed a framework in which it is possible to describe the Chameleon-like screening mechanisms with two free functions. \citet{2015arXiv150507129G} present a fully empirical parametrization of screened modified gravity models using three parameters which captures a number of models \& model parameters.

In this paper, we want to revisit some classical quantities associated with screened modified gravity models, namely the dynamical, lensing and thermal mass estimates of clusters of galaxies in the light of \textit{(i)} the \citet{2015arXiv150507129G} parameterisation, and, \textit{(ii)} the possibility of unequal coupling; that the enhancement of gravity is not the same for baryons and dark matter. 

In this work, we use $M_{\rm Pl}^{-2} \equiv 8 \pi G$, $\rho_c = 3 H^2 M_{\rm Pl}^2$, and, denote values today with a subscript zero.

\section{Methods}
\label{sec:methods}

\begin{figure*}
  \includegraphics[width=.49\textwidth]{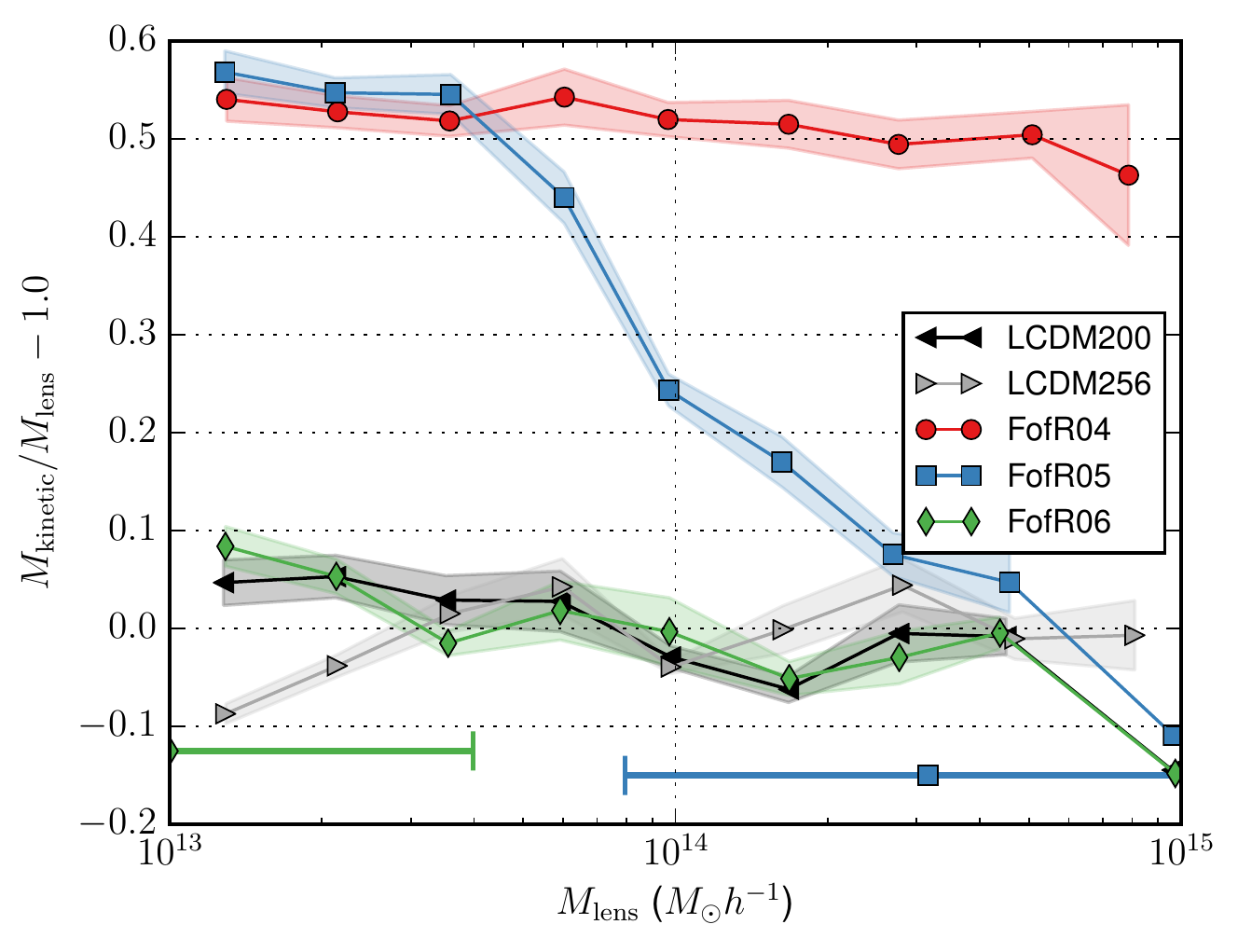}
  \includegraphics[width=.49\textwidth]{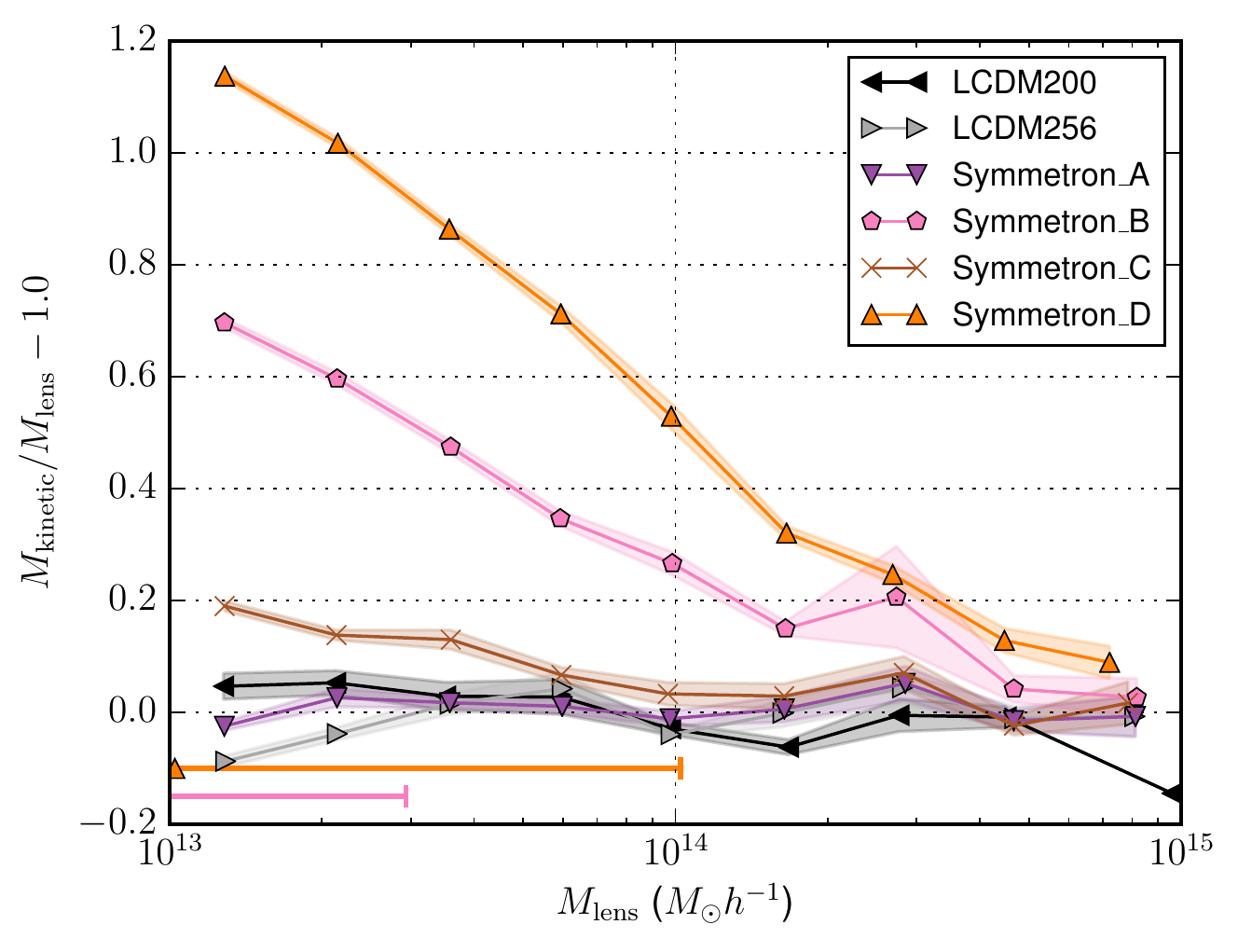}
  \caption{Ratio of kinetic and lensing mass for the analyzed $f(R)$ models (left panel) and Symmetron models (right panel). The horizontal lines and markers show the width and center of the transition region as defined in \S~\ref{sec:univ-param-scre}, respectively.}
  \label{fig:Mkin}
\end{figure*}

\begin{figure*}
  \includegraphics[width=.49\textwidth]{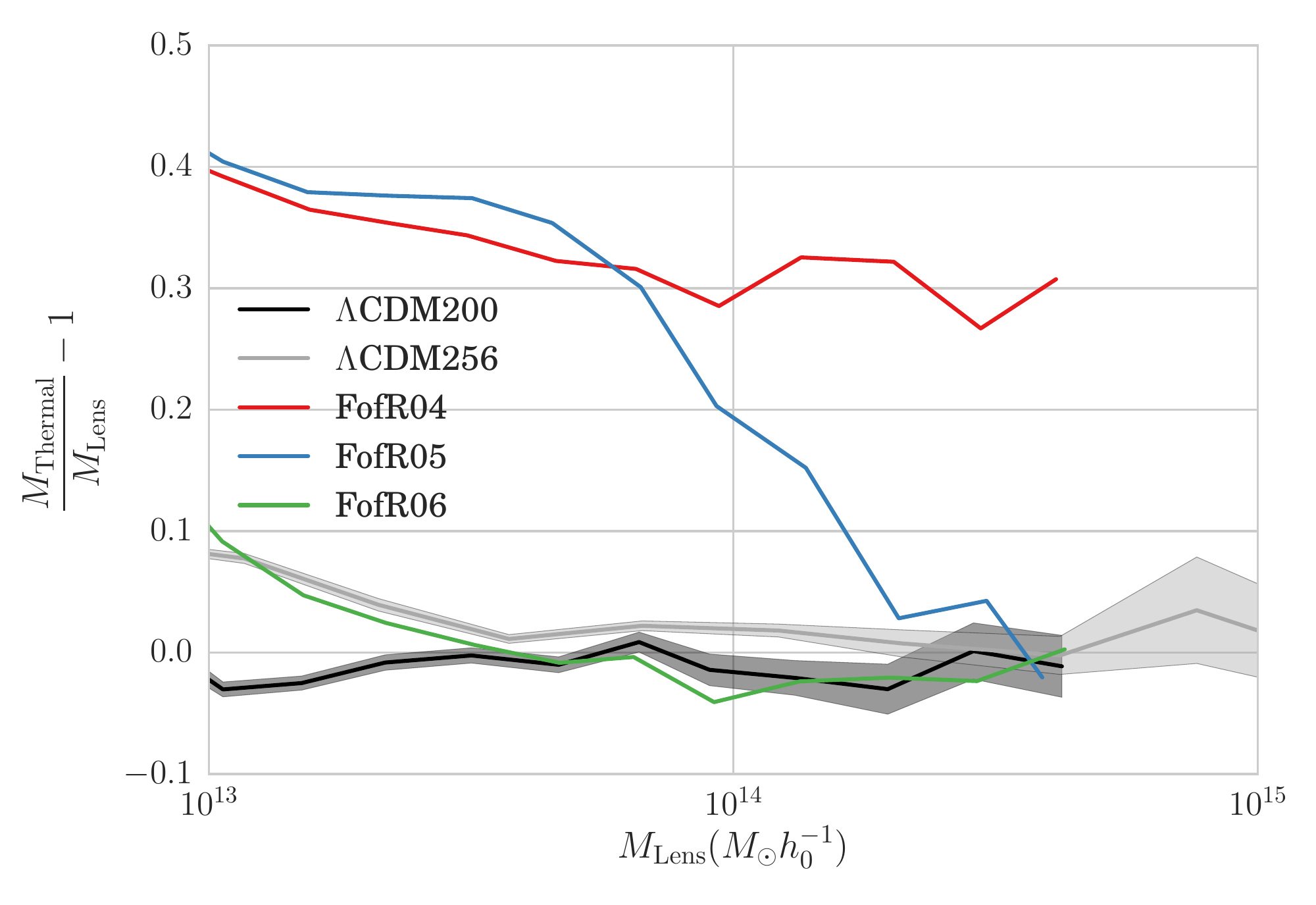}
  \includegraphics[width=.49\textwidth]{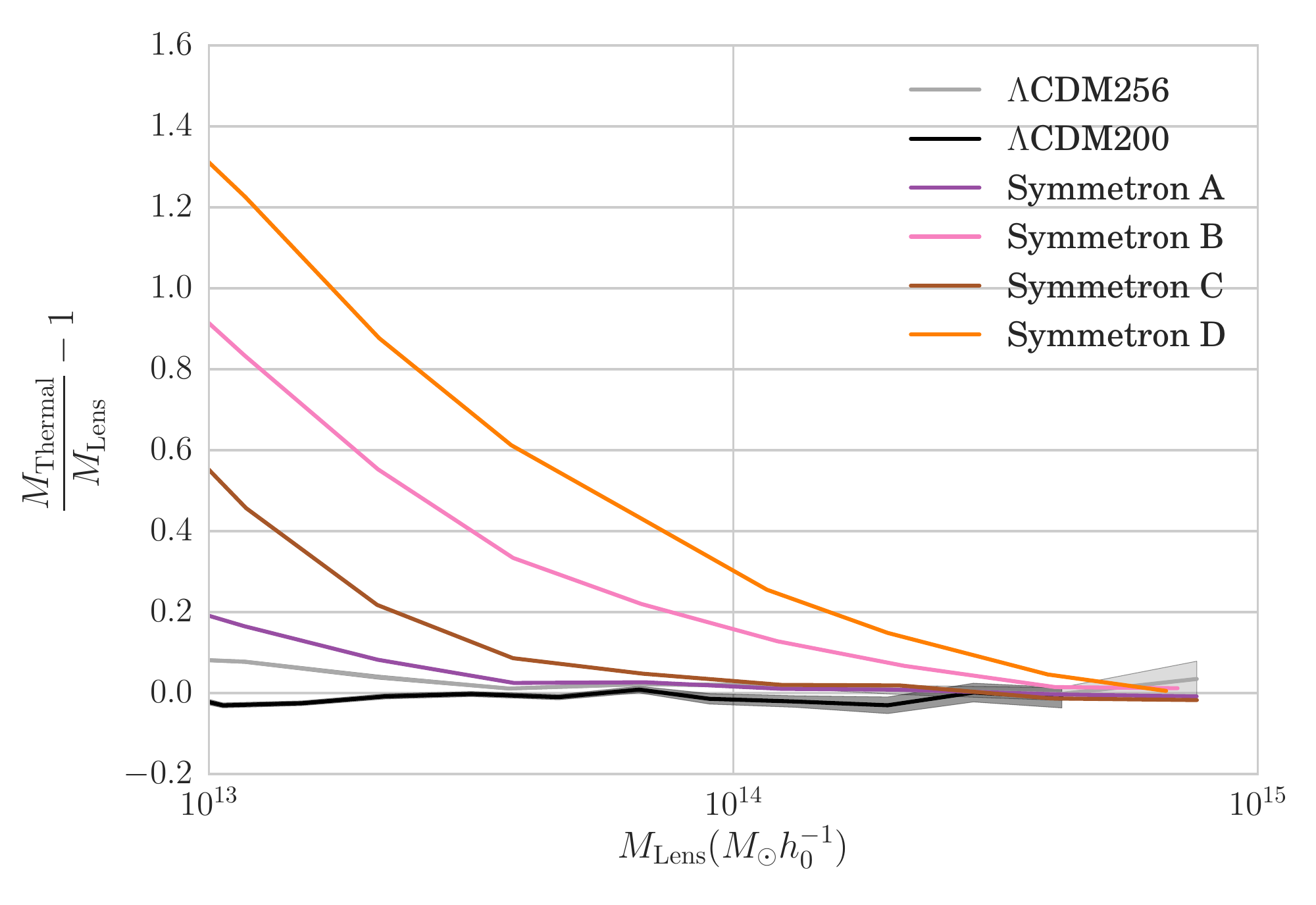}
  \caption{Ratio of thermal and lensing mass for the analyzed $f(R)$ models (left panel) and Symmetron models (right panel). The horizontal lines and markers show the width and center of the transition region as defined in \S~\ref{sec:univ-param-scre}, respectively.}
  \label{fig:Mtherm}
\end{figure*}

\subsection{Screened modified gravity models}
In this subsection we introduce very briefly the Symmetron \citep{Hinterbichler,Hinterbichlera} and the \citet{Hu2007} $f(R)$ model. For more details we refer to the original papers, to reviews featuring these models \citep[e.g.,][]{Clifton2012,2013arXiv1312.2006K}, or, to our previous work where we introduce the models in more detail \citep[e.g.][]{Gronke2014_dl,Hammami2015a}.

The \citet*{Hu2007} $f(R)$ model is a $f(R)$ model featuring the Chameleon screening mechanism, i.e., it has a reduction of both the range and the strength of the fifth force in high density regions. The model features two free parameters, $|f_{R0}|$ and $n$, where former controls the range of the fifth force in vacuum and the latter does not have much impact \citep{Hu2007}. We consider three models with $\log_{10}|f_{R0}| = (-4,\,-5,\,-6)$ and $n=1$. For the \citet*{Hu2007} $f(R)$ model, the maximum enhancement of the gravitational force with respect to GR is fixed to $\gamma_{\rm max}^{f(R)} = 1/3$.

The Symmetron \citep{Hinterbichler,Hinterbichlera} inherits a symmetry breaking effective potential leading to a diminishing fifth-force in high-density environments. The model parameters are the scale factor of average symmetry breaking $a_{\rm ssb}$, the range of the force in vacuum $\lambda_\psi$, and, the coupling parameter $\beta$. The maximum enhancement of gravity is in this case \citep{2015arXiv150507129G}
\begin{equation}
\gamma_{\rm max}^{\rm Symmetron} = 2 \beta^2 \left[ 1 - \left(\frac{a_{\rm ssb}}{a}\right)^3\right]\;.
\label{eq:gammamax_symm}
\end{equation}

\subsection{$N$-body simulations \& halo selection}
We used a modified version \citep{Hammami2015a} of the \texttt{ISIS} \citep{Llinares2014a} simulation, which in turn is based on the adaptive-mesh code \texttt{RAMSES} \citep{2002A&A...385..337T}. The initial conditions were created using Grafic \citep{Grafic} starting at a redshift $z = 49$. The simulation parameters used are $(\Omega_{CDM0},\,\Omega_{b0},\,\Omega_{\Lambda0},\,H_0,\,B,\,N)=(0.227, 0.045, 0.727, 70\kms\,{\rm Mpc}^{-1}, 200\Mpch, 256)$ for the $f(R)$ simulation set and $(0.3, 0.05, 0.65, 65\kms\,{\rm Mpc}^{-1}, 256\Mpch, 256)$ for the Symmetron simulation set, where $B$ denotes the side-length of the simulation box and $N$ the number of particles in the box. Furthermore, the modified gravity parameters used are identical to \citet{Hammami2015a,Hammami2015b} and are summarized in Table~\ref{tab:mgparam1} and Table~\ref{tab:mgparam2}. The simulations contains both dark matter particles and baryons which are treated as a simple ideal fluid\footnote{This means that no additional baryonic-physics is included in our simulations like star-formation, cooling, feedback etc.}. All the results in this paper comes from analyzing the $z=0$ snapshot of the simulations.

\begin{table}
 \begin{center}
\caption{Overview of the \textit{equal coupled} model parameters for the Symmetron and $f(R)$ models.}
\label{tab:mgparam1}
  \begin{tabular}{lrrr}\hline
  Symmetron models  & $\beta$ & $ a_{\rm SSB}$ & $\lambda_{\psi}$ \\ \hline
  \symmA & 1.0 & 0.5 & 1.0 \\
  \symmB & 1.0 & 0.33 & 1.0 \\
  \symmC & 2.0 & 0.5 & 1.0 \\ 
  \symmD & 1.0 & 0.25 & 1.0 \\ \hline\hline\\\hline
  $f(R)$ models   & $ f_{R0}$&$ n$  \\ \hline
  \fofrfour & $10^{-4}$&1& \\ 
  \fofrfive & $10^{-5}$&1& \\ 
  \fofrsix & $10^{-6}$&1& \\ \hline\hline
  \end{tabular}
 \end{center}
\end{table}

\begin{table}
  \centering
  \caption{Overview of the \textit{mixed coupled} models. All other parameters are identical to the `\symmB' model (see Table\ref{tab:mgparam1}).}
  \begin{tabular}{lrrr}\hline
  Unequal coupled models & $\beta_{\rm DM}$ & $\beta_{\rm Gas}$ \\ \hline
  \BothOne & 1.0 & 1.0\\
  \BothPointOne & 0.1 & 0.1\\
  \DMPointOne & 0.1 & 1.0\\
  \GasPointOne & 1.0 & 0.1\\
\end{tabular}
\label{tab:mgparam2}
\end{table}

The halos were identified using the spherical overdensity halo finder \texttt{AHF} \citep[Amiga Halo Finder,][]{AHF}. For the analysis we used only halos consisting of at least $100$ particles which limits the smallest halo we can probe to $M\sim 3\times 10^{12}M_{\odot}h^{-1}$. In the high mass end the simulation-box limits the maximum halo-masses we can study and the largest halos in our simulations has mass $M \sim 2-3\times 10^{15} M_{\odot}h^{-1}$. We have checked that the mass-function of our simulation agrees to $\sim 10\%-20\%$ to simulations with larger box-size and also to the \citet{2008ApJ...688..709T} fit to the mass-function in the range $M\in [ 10^{13},8\times 10^{14}] M_{\odot}h^{-1}$. The total number of halos in this mass-range in our simulations, which is what we used for the upcoming analysis, is $\sim 8000$.

\subsection{Halo mass estimates}
\label{sec:halo-mass-estimates}
After identifying the halos with \texttt{AHF} we define three kinds of mass estimates:

\begin{enumerate}
\item The \textit{lensing mass} $M_{\rm lens}$ as the $M_{200c}$ as given by \texttt{AHF}.
\item The \textit{thermal mass} $M_{\rm therm}$ constructed using the temperature- and density profiles as
\begin{equation}
M_{\rm therm} = -\frac{k_Br^2T_{\rm thermal}(r)}{\mu m_p G}\left(\frac{d\ln\rho_{\rm thermal}}{dr} + \frac{d\ln T_{\rm thermal}}{dr}\right),
\end{equation}
where $k_B$ is the Boltzmann constant, $r \sim R_{200c}$ is the virial radius of the halo, $m_p$ is the proton mass and $\mu=0.59$ is the mean molecular weight of the gas\footnote{Note, that we did not include the non-thermal pressure component here to match observational procedure. See \S\ref{sec:therm-mass-meas} and Appendix \ref{sec:appendix} for a discussion of the non-thermal contribution and its implications.} and,
\item the \textit{kinetic mass} $M_{\rm kin}$ calculated from the velocity disperion via
\begin{equation}
M_{\rm kin} = M_0 \left(\frac{\sigma_{\rm DM}}{\sigma_0}\right)^{1/\alpha}
\label{eq:mkin}
\end{equation}
where $M_0$, $\sigma_0$ and $\alpha$ are fitting values, and, $\sigma_{\rm DM}$ is the one-dimensional velocity dispersion of all the dark matter particles within the virial radius $R_{200c}$. For $(M_0,\,\sigma_0,\,\alpha)$ we adopt the values found by \citet{Evrard2008}, namely $(M_0,\,\sigma_0\,\alpha) = (10^{15}\Msh,\,1082.9\kms,\,0.3361)$.
\end{enumerate}

\subsection{A universal parametrization of screened modified gravity models}
\label{sec:univ-param-scre}
The parametrization of \citet{2015arXiv150507129G} is based on the simple idea that in screened modified gravity theories, the fifth force is screened for the most massive halos, while being completely unscreened for the very lightest halos and voids, where the fifth force is in full play and achieves it's theoretical maximum value. This implies that there has to be a transition scale where the enhancement of the gravitational force is about half its theoretical maximum.

In \citet{2015arXiv150507129G}, this transition scale is phrased in terms of halo mass ($M_{200}$), and thus, dubbed $\mu_{200}$. In other words: The (mass weighted) average fifth force in a halo with mass $\mu_{200}$ is roughly $\gamma_{\rm max}/2$ that of the Newtonian prediction. The second parameter is the width of this transition region, say, when the enhancement of gravity is between $20$ and $80$ percent of $\gamma_{\rm max}$. This width is quantified with a third half-width parameter $W$. In conclusion, this means that fully screened (completely unscreened) halos are expected to have masses $\lesssim \mu_{200} / W$ ($\gtrsim \mu_{200} W$) and, thus, the effective gravitational constant in these halos is simply $G$ ($\gamma_{\rm max} G$).

In order to map the parameters of several screened-modified gravity theories \citet{2015arXiv150507129G} solved the full field equations on an isolated NFW density profile obtaining full gravitational force profiles -- and hence also the mass averaged enhancement of gravity -- for a wide range of model parameters and halo masses. Additionally, they also use the more general screened modified description of \citet{2012PhRvD..86d4015B} to obtain a similar semi-numerical mapping. As a result \citet{2015arXiv150507129G} provide fitting formulas for the DGP, Hu-Sawicki $f(R)$ and Symmetron model parameters to the universal $(\mu_{200},\,\gamma_{\rm max},\,W)$ set of which we use the latter two in \S\ref{sec:observ-constr}. For more details on how these re-mapping rules were developed we refer the reader to \citet{2015arXiv150507129G}.

\begin{figure*}
  \includegraphics[width=.49\textwidth]{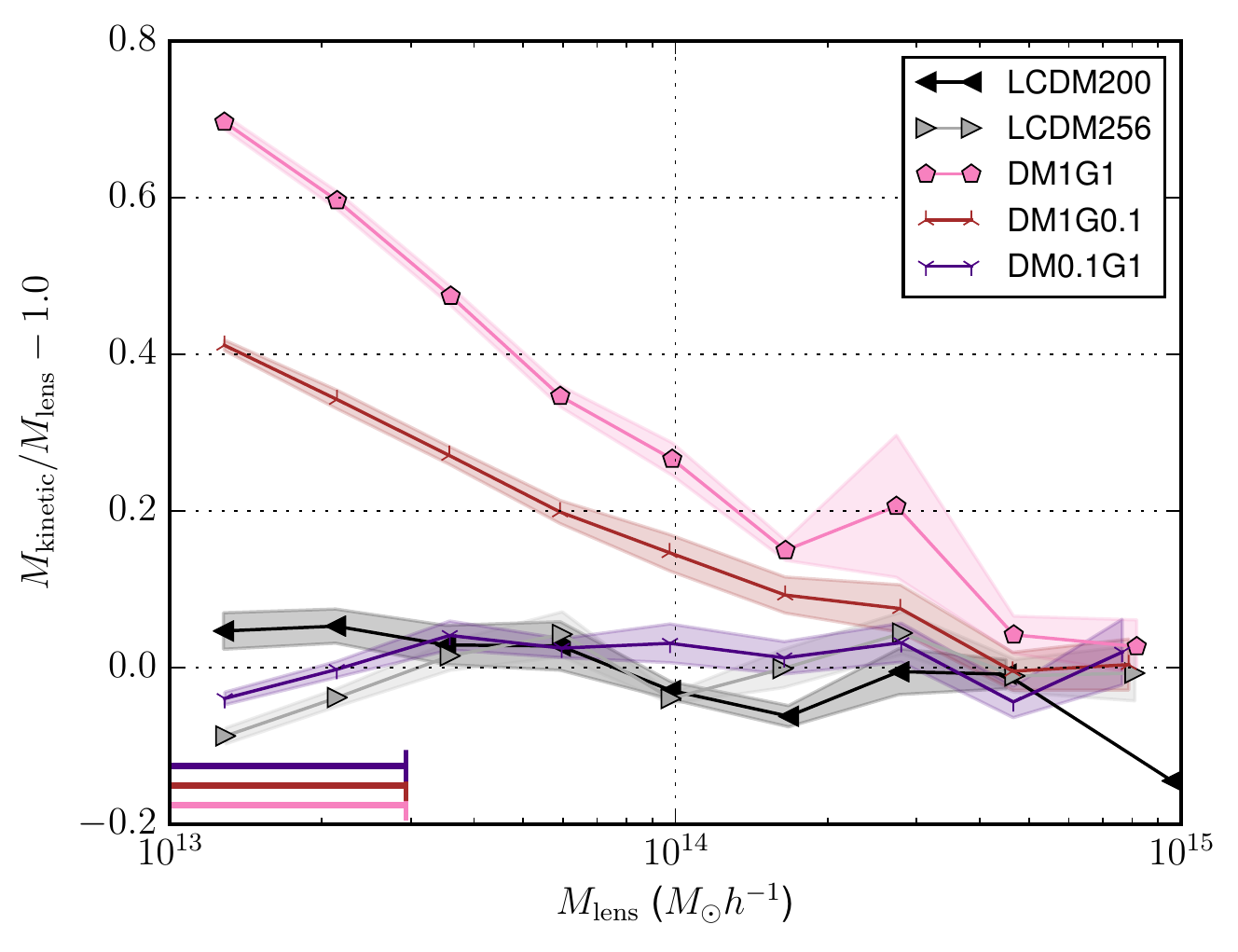}
 \includegraphics[width=.49\textwidth]{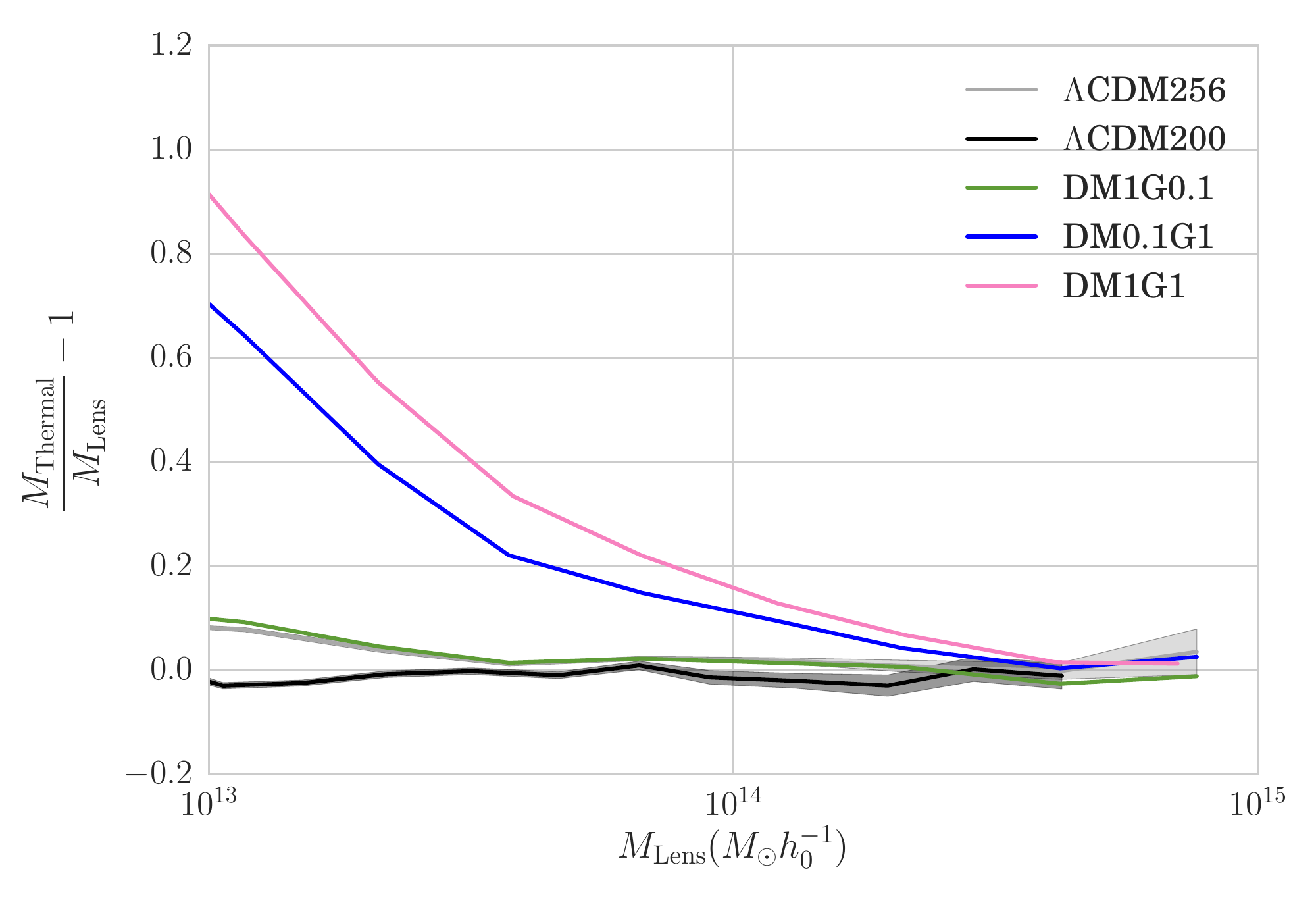}
  \caption{Mass ratios for the analyzed models with non-universal coupling. The \textit{left panel} shows the ratio of the kinetic- and lensing mass and the \textit{right panel} shows the ratio of the thermal and lensing mass.}
  \label{fig:mixed}
\end{figure*}

\begin{figure}
  \includegraphics[width=.49\textwidth]{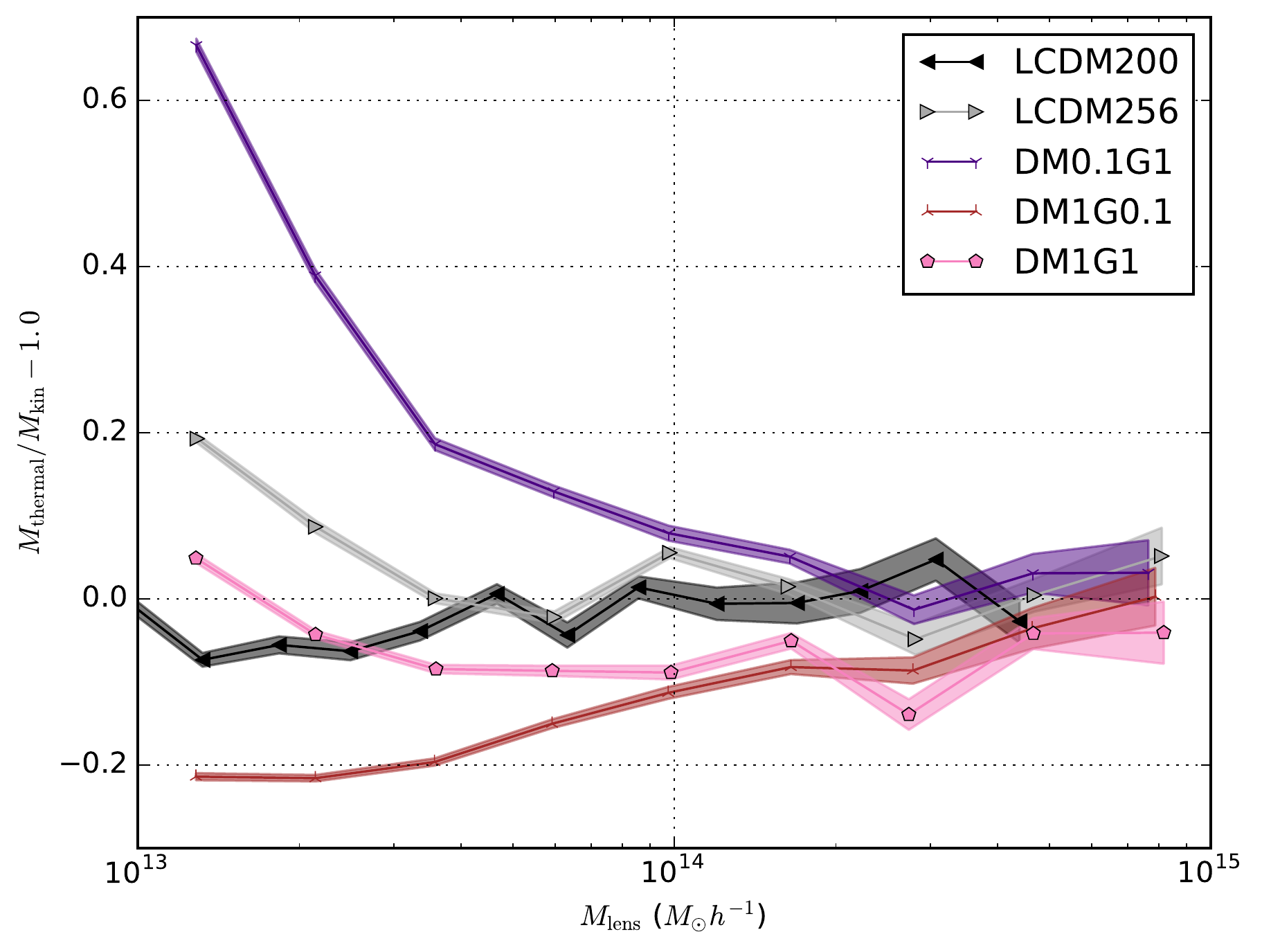}
  \caption{Mass ratios between the thermal and kinematic mass for the models with non-universal coupling (see \S~\ref{sec:results-non-universal}).}
  \label{fig:mixed2}
\end{figure}

\section{Results}
\label{sec:results}
In this section we present the results from the numerical $N$-body simulation in the case of universal (\S~\ref{sec:result-universal-coupling}) and non-universal coupling (\S~\ref{sec:results-non-universal}).

\subsection{Models with universal coupling}
\label{sec:result-universal-coupling}

Fig.~\ref{fig:Mkin} shows the ratio of the lensing and kinetic masses (as given by Eq.~\ref{eq:mkin}) for the simulated $f(R)$ (left panel) and the Symmetron models (right panel). The shaded bands in this figure denote the error of the mean in each bin. Similarly, Fig.~\ref{fig:Mtherm} shows the ratio of the thermal mass and the lensing mass for the analyzed models.

Both figures show the same -- well known -- features: \textit{(i)} a large deviation for smaller mass halos (which are unscreened); \textit{(ii)} a decline for intermediate masses when the screening kicks in; and, \textit{(iii)} the fully screened high-mass halos where the deviation is essentially zero. We want to highlight, however, that this deviation is between measures of the same simulation. This is in contrast to what is often presented in similar studies -- where the deviation between the modified gravity and the $\Lambda$CDM simulation is displayed. Therefore, finding similar trends as these studies \citep[e.g.][]{2014MNRAS.440..833A,Gronke2014_dl,Falck2015JCAP...07..049F} is reassuring.

In addition to the data points, Fig.~\ref{fig:Mkin} also displays the estimated transition scales (as described in \S\ref{sec:univ-param-scre}). Specifically, the value of expected centroid of the transition scale $\mu_{200}$ and the half-width of the transition $W$ is marked with a matching symbol and colored lines, respectively.

\subsection{Models with non-universal coupling}
\label{sec:results-non-universal}

Fig. \ref{fig:mixed} shows the mass ratios $M_{\rm kinetic} / M_{\rm lens}$ and $M_{\rm thermal} / M_{\rm lens}$ (left and right panel, respectively) for the models with non-universal coupling, i.e., for which the fifth force acts differently on the baryons and the dark-matter. The models presented are variations of the \symmB model (see Table~\ref{tab:mgparam2}). Clearly, the same trends as in \S~\ref{sec:result-universal-coupling} are visible. However, this time different coupling combinations are sensitive to different observables. In particular, the model where dark matter is stronger coupled is more sensitive to the kinetic mass estimate, and, the model where the baryons are stronger coupled shows a (much) stronger variation in the thermal mass.

This effect can be seen more clearly in Fig.~\ref{fig:mixed2} where we show the ratio between the thermal and the kinetic mass. Here, the model with stronger baryonic coupling shows a clear positive deviation $\gtrsim 50\%$, and the stronger dark matter coupled model a negative deviation. This is interesting as the equally coupled model is much closer to the $\Lambda$CDM prediction. Note, that the discrepancy between the two mass scales in both the $\Lambda$CDM cases comes from the imperfect calibration of our mass-estimates and the resulting small deviations at the low-mass end. This could be overcome using higher-resolution simulations or better calibrated predictors.

This means that by studying solely the $M_x / M_{\rm lens}$ ratios in Fig.~\ref{fig:mixed} one could construct a model which mimics the effect of a universal coupling (or vice versa). For instance, increasing the dark-matter coupling to $\beta_{\rm DM}\sim 1.0$ (while leaving $\beta_{\rm gas}=0.1$ untouched) one could obtain a similar kinematic mass estimate as in the \symmB model. Or, to state another example, increasing the coupling to baryons for the `DM0.1G1' model slightly will lead to an thermal mass estimate as found in \symmB. However, this degeneracy can be broken when comparing directly the thermal, and, kinetic mass estimates as in Fig.~\ref{fig:mixed2}.

\begin{figure}
  \includegraphics[width=.49\textwidth]{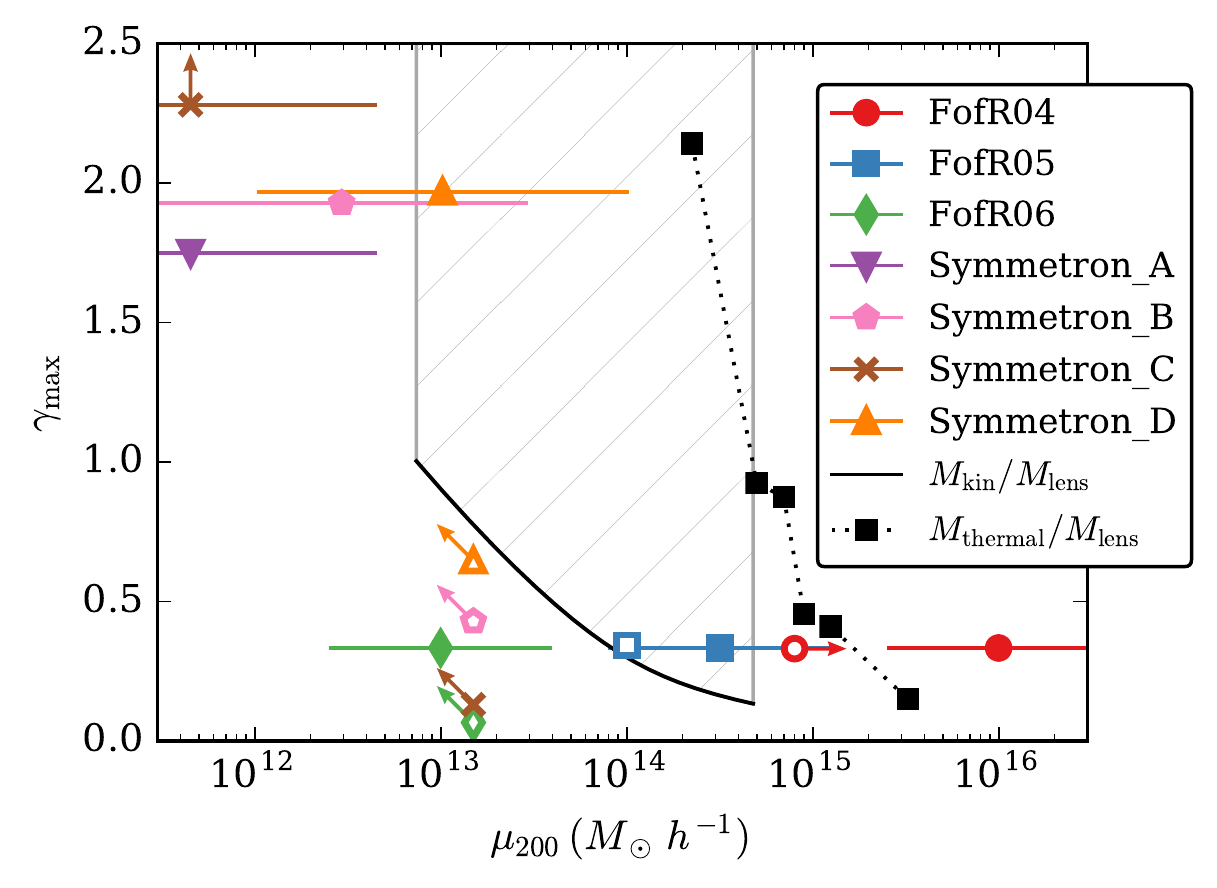}
  \caption{Observational constraints on screened modified gravity theories presented in the $\mu_{200}-\gamma_{\rm max}$ parametrization described in \S\ref{sec:univ-param-scre}. The filled symbols with horizontal lines show the semi-analytic $\mu_{200}$ and $W$ predictions. The unfilled symbols show the simulation results with arrows denoting limits due to the mass resolution of the simulations. The black line shows the constraints from $M_{\rm kin}$ and $M_{\rm lens}$ observations resulting in the exclusion of the grey shaded region (see \S\ref{sec:observ-constr} for details).}
  \label{fig:obs_constraints}
\end{figure}

\begin{figure*}
  \includegraphics[width=.9\textwidth]{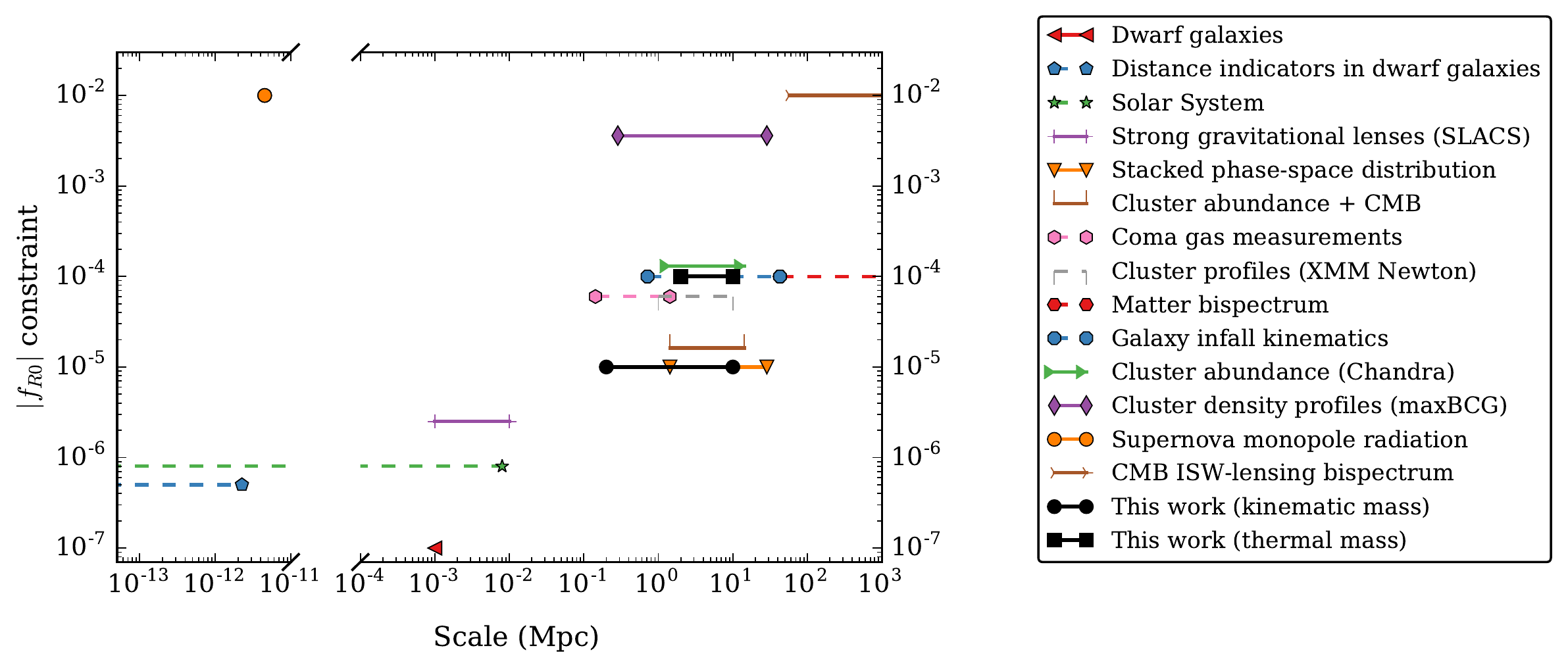}
  \caption{Current constraints on the Hu-Sawicki $f(R)$ model with $n=1$. References \citep[updated list from][]{2014AnP...526..259L}: Dwarf galaxies \citep{jain:11,vikram:13}, Distance indicators in dwarf galaxies \citep{jain:13}, Solar System \citep{Hu2007,lombriser:14}, Strong gravitational lenses (SLACS) \citep{Smith2009arXiv0907.4829S}, Stacked phase-space distribution \citep{2012PhRvL.109e1301L}, Cluster abundance + CMB \citep{Cataneo2015PhRvD..92d4009C}, Coma gas measurements \citep{2014JCAP...04..013T}, Cluster profiles (XMM Newton) \citep{2015MNRAS.452.1171W}, Matter bispectrum \citep{2011JCAP...11..019G}, Galaxy infall kinematics \citep{2014MNRAS.445.1885Z}, Cluster abundance (Chandra) \citep{2009PhRvD..80h3505S,2011PhRvD..83f3503F}, Cluster density profiles (maxBCG) \citep{2012PhRvD..85j2001L}, Supernova monopole radiation \citep{2013arXiv1306.6113U}, CMB ISW-lensing bispectrum \citep{2013PhRvD..88b4012H,2014MNRAS.442..821M}.}
  \label{fig:obs_constraints_fofr}
\end{figure*}

\section{Discussion}

\subsection{Comparison to semi-analytic predictions}
\label{sec:comp-semi-analyt}

The semi-analytic predictions of \citet{2015arXiv150507129G} for the position and width of the transition scale as well as the maximum enhancement of gravity (see \S~\ref{sec:univ-param-scre}) can be tested against the results from the $N$-body simulation.

Our findings can be summarized as follows.
\begin{itemize}
\item The centroid of the transition region, $\mu_{200}$, is lower in the $N$-body simulations than predicted by the semi-analytic model. This is clearly visible for the \fofrfive model, as here the full transition is within the mass scale of our simulation. In this case the offset is circa half an order of magnitude. This offset is due to the fact that in \citet{2015arXiv150507129G} the impact of the fifth force on a complete, isolated halo -- i.e., out to $R=10R_{\rm vir}$ -- was analyzed. As in reality, e.g., tidal effects from nearby halos play an important role for the behavior in the outer regions, we constraint our analysis here to $1 R_{\rm vir}$ -- which is also closer to observations. Other, however, sub-dominant factors are the inclusion of baryonic effects in this study \citep[see][for a full discussion of baryonic effects on modified gravity simulations]{Hammami2015a}, and, the higher environmental density of the halos in the $N$-body simulation.
\item The half-width of the transition region, $W$, can only be compared to the \fofrfive model for the reasons explained above. In this one case, the prediction fit quite well.
\item For a virialized halo, the velocity dispersion squared is proportional to the gravitational constant -- as per definition a in a virialized halo two times the kinetic is equal to minus the potential energy. Using now the definitions of the velocity dispersion (Eq.~\eqref{eq:mkin}), of $\gamma$, and requires that $M_{\rm kin}\rightarrow M_{\rm lens}$ for $\gamma\rightarrow 0$, one finds
\begin{equation}
M_{\rm kin} / M_{\rm lens} = (\gamma + 1)^{1/2\alpha}\;. 
\label{eq:Mkin_Mlens_ratio}
\end{equation}
Thus, the maximum deviation found for this mass ratio for the $f(R)$ models (see Fig.~\ref{fig:Mkin}) fits the theoretical estimate of $\sim 0.53$ using $\gamma_{\rm max}=1/3$ (and $\alpha = 0.3361$ as described in \S\ref{sec:halo-mass-estimates}). For the Symmetron models considered, the unscreened masses lie below the mass resolution of our simulations making the prediction untestable.
\item Likewise we can construct a relation for the thermal mass. Using the relation of \citet{2014JCAP...04..013T} $M = M_{GR} + M_{F_{\psi}}$, we obtain $M_{\rm thermal} / M_{\rm lens}\propto \gamma + 1$. The maximum deviation found for the thermal mass ratio for the $f(R)$ models can then be read from Fig.~\ref{fig:Mtherm}, and is consistent with the theoretical prediction $\gamma_{\rm max}=1/3$.
\end{itemize}

Fig.~\ref{fig:obs_constraints} shows the $\mu_{200}-\gamma_{\rm max}$ predictions (filled symbols) with the width of the transition region marked as horizontal line. The same figure displays the resulting parameters from the $N$-body simulation as unfilled symbols with arrows denoting limits due to the mass resolution of the simulations.

\subsection{Observational constraints}
\label{sec:observ-constr}
Masses of clusters can be inferred using different mass-estimates. Within conformally invariant modified gravity theories with a screening mechanisms, different mass-estimates may result into different inferred values for the mass. For instance, while the mass inferred via lensing gives the same value as in General Relativity (independently of the size and the environment of the halo), the mass measured via dynamical methods (e.g. inferred from velocity dispersion measurements) may result into a different value, specially for small and isolated halos \citep{2012ApJ...756..166W}. In this section, we use existing lensing and dynamical mass measurements and compare them with each other in order to constrain the modified gravity parameter space. 

\subsubsection{Lensing versus kinetic mass measurements}
We use lensing and kinematic mass measurements from the Sloan Digital Sky Survey \citep[SDSS][]{2009ApJ...703.2217S}. In particular, we use the lensing and kinetic mass estimates from \citet{arXiv0709.1159J} and \citet{2007ApJ...669..905B}, respectively. In order to combine the two masses, we use the richness of a cluster $\tilde N_{200}$ which represents the number of detected galaxies associated with a cluster\footnote{$\tilde N_{200}$ is dependent on the limiting magnitude of the survey. Therefore, we ensured that the our mass measurement data both used the Sloan Digital Sky Survey \citep[SDSS][]{2009ApJ...703.2217S} $\tilde N_{200}$ data} and is therefore independent of the chosen mass estimate. This procedure allows us to constrain the maximally allowed deviation from GR across several mass-scales. 

For the lensing masses we use the mass-richness relation given by \citet{arXiv0709.1159J} who used $130,000$ clusters of galaxies. They found a relation given by
\begin{equation}
\tilde M_{\rm lens} = (8.8 \pm 1.17)\times 10^{13} \left(\frac{\tilde N_{200}}{20}\right)^{1.28\pm0.04} M_\sun h^{-1}
\end{equation}
where $\tilde N_{200}$ is again the measured richness of the cluster.

The dynamical mass measurements were taken from \citet{2007ApJ...669..905B} who found
\begin{equation}
b_v \tilde M_{\rm kin} = (1.18\pm 0.12)\times 10^{14}\left(\frac{\tilde N_{200}}{25}\right)^{1.15\pm 0.12}
\end{equation}
where $b_v$ denotes the bias.

When combining the observations we fixed the value of the bias $b_v$ to its maximum value under the constraint that the two mass measurements agree within $1\sigma$ throughout the considered mass-range -- i.e., in the range $M / M_\sun h^{-1}\in [7\times 10^{12},\,5\times 10^{14}]$ which is the overlapping mass-range of the observations -- yielding $b_v = 1.03$.
This leaves us with a conservative constraint for the maximally allowed over-prediction of the kinetic mass compared to the lensing mass. This constrained can be converted to a maximally allowed enhancement of gravity using Eq.~\eqref{eq:Mkin_Mlens_ratio} for every halo-mass in the considered mass range, i.e., we can map the observational limits of $M_{\rm kin} / M_{\rm lens}$ to an maximally allowed $\gamma(M)$ -- or, in the language of the universal description of \citet{2015arXiv150507129G} introduced in \S\ref{sec:univ-param-scre} to a constraint in the $(\gamma_{\rm max},\,\mu_{200})$-plane. To recap: In this picture, $\mu_{200}$ is the halo mass where the enhancement of gravity (and, thus, the ratio $M_{\rm kin} / M_{\rm lens}$ or $M_{\rm therm}/M_{\rm lens}$ reaches half of its theoretical value, and $\gamma_{\rm max}$ is this maximally enhancement of gravity (and, therefore, sets via Eq.~\eqref{eq:Mkin_Mlens_ratio} the upper bound on the mass ratio).

Fig.~\ref{fig:obs_constraints} shows the $2-\sigma$ observational constraint as a black line, and the resulting ruled out region of the $\gamma-\mu_{200}$ parameter space as grey shaded region. Note, that this region does not extent to greater masses as one might naively assume as higher kinetic masses throughout the entire probed mass-range can be explained with a constant bias. However, these greater masses are ruled out by halo abundance measurements for $\gamma_{\rm max} \sim \mathcal{O}(1)$ \citep[e.g.][]{Cataneo2015PhRvD..92d4009C}.

In addition, Fig.~\ref{fig:obs_constraints_fofr} shows this constraint on the model-dependent parameter space for the Hu-Sawicki $f(R)$ model with $n=1$ which we obtained by comparing our full simulation results (Fig.~\ref{fig:Mkin}) to the allowed mass ratio deviation in each mass bin. This means we did not use the semi-analytical $(\mu_{200},\,\gamma_{\rm max})-(M_{\rm kin},\,M_{\rm therm})$-relation but used the full simulation and, hence, could rule out only some few values of $|f_{R0}|$. In spite of that this rather conservative limit\footnote{In contrast to other studies, we \textit{(i)} used a full $N$-body simulation with baryonic effects as calibration, and, \textit{(ii)} assumed a `worst-case' bias as stated above.}, using the kinematic mass estimates of clusters is still competitive with other measurements at this length-scale. 

\subsubsection{Thermal versus lensing mass measurements}
\label{sec:therm-mass-meas}

In order to compare the lensing and thermal mass of the clusters we took measurements from \citet{2010ApJ...711.1033Z} and \citet{2013ApJ...767..116M}. These two data-sets provide both thermal and lensing mass measurements and uncertainties for a total of $58$ clusters in the mass range $M/M_{\sun} h^{-1}\in[5\times 10^{13},\,3\times 10^{15}]$ so there was no need to combine the mass estimates in a similar fashion as in the previous section. We divided the data for the thermal mass measurements by the data for the lensing mass measurements while properly propagating the error. As we're interested in a systematic deviation, we binned the data in $6$ (lensing) mass bins which we stratified so that roughly the same number of halos are in each bin.  

An important point to bear in mind when working with thermal mass estimates is the fact that the measured quantity in this case is the temperature of the intracluster gas. The conversion to a mass assumes hydrostatic-equilibrium \citep[as done, e.g., in][]{2010ApJ...711.1033Z,2013ApJ...767..116M,2014MNRAS.440..833A}. However, it has been shown that in reality the pressure of the intracluster medium will have a significant non-thermal component generated by random gas motions and turbulence \citep{Evrard1990,Rasia2004,Kay2004,Dolag2005,Lau2009}.
This means the inferred thermal mass given a temperature $T$ will be slightly lower than the true mass of the cluster. 

While empirical models exist in order to quantify the magnitude of this deviation (where the non-thermal component yield variations to the mass from 10\% to 30\%  \citep{2010A&A...510A..76L}) we want to stress that these were calibrated against pure $\Lambda$CDM simulations, and thus their results cannot be taken into account when dealing with modified gravity.
One has to consider instead that if gravity is truly enhanced, the temperature of the intra-cluster medium will be hotter and, thus, the inferred thermal mass will be greater (as shown in Sec.~\ref{sec:results}).
This means the effect of any non-thermal physics \citep[such as cosmic rays][]{2008MNRAS.385.1242P} is degenerate with modified gravity and, consequently, at the present time thermal measurements cannot be used to constrain modified gravity. 

\begin{figure}
  \includegraphics[width=.49\textwidth]{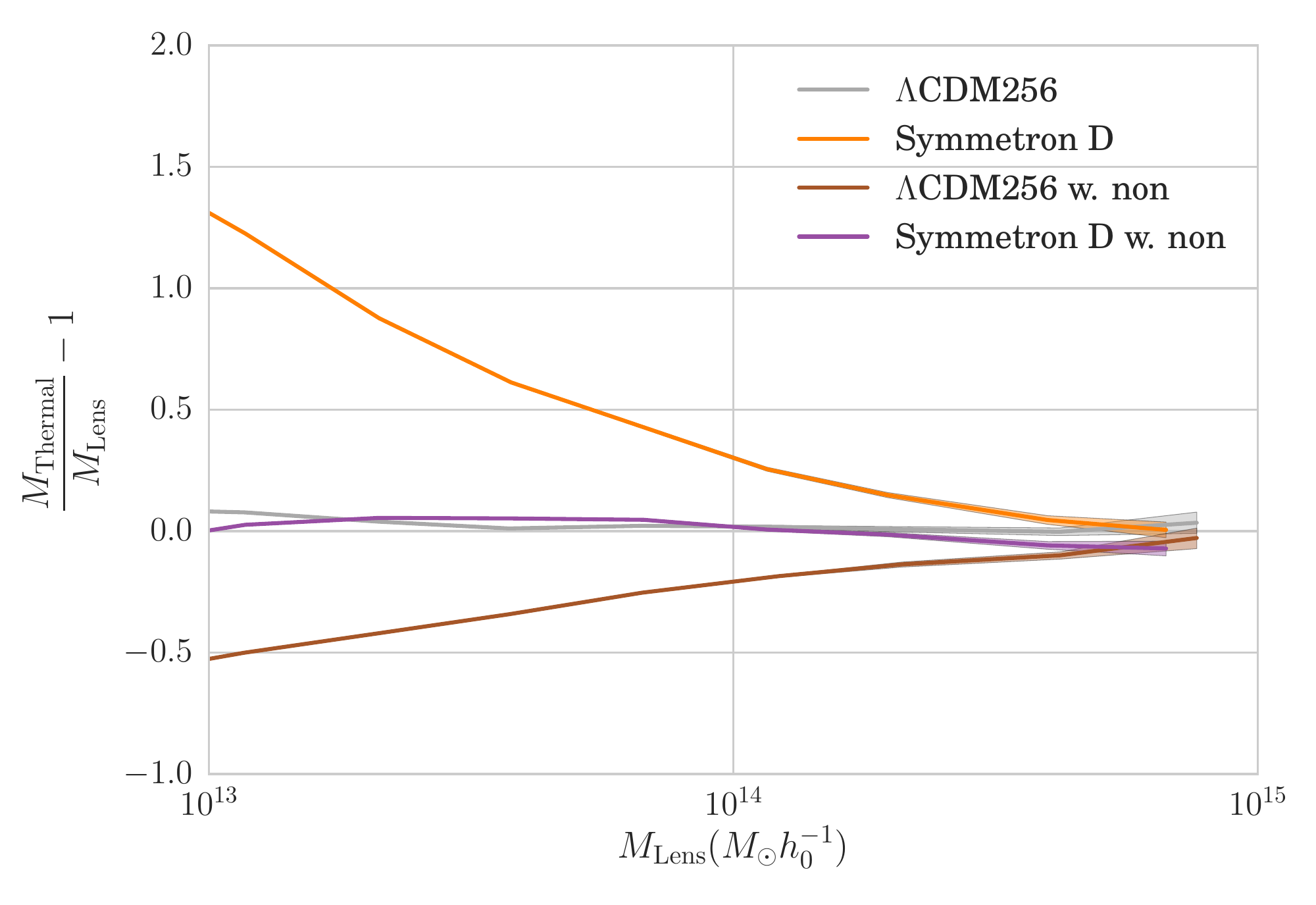}
  \caption{Ratio of thermal and lensing mass for $\Lambda$CDM and Symmetron D, the symmetron model with the largest deviations from $\Lambda$CDM. Also included is the hypothetical measurements obtained by including the non-thermal contribution described in \S~\ref{sec:therm-mass-meas}}
  \label{fig:non-thermal-degen}
\end{figure}

Fig.~\ref{fig:non-thermal-degen} illustrates this degeneracy. Here, we compare the $M_{\rm therm}/M_{\rm lens}$ results of the Symmetron D model and our $\Lambda$CDM simulation (already shown in Fig.~\ref{fig:Mtherm}) to hypothetical measurements where we modeled the contribution of non-thermal pressure as 
\begin{equation}
P_{\rm non-thermal} = P_{\rm total} \tilde g(M_{\rm 200}) 
\label{eq:Pnontherm1}
\end{equation}
which resembles the functional forms fitted to $\Lambda$CDM simulations (\citet{Shaw2010, Battaglia2012} and see also Appendix~\ref{sec:appendix}). Thus, our proposed non-thermal contribution is not unreasonable. Modeling a non-thermal contribution as given by Eq.~\eqref{eq:Pnontherm1} while keeping the total pressure $P_{\rm total} = P_{\rm therm} + P_{\rm non-thermal}$ (and, thus, the halo structure) constant is equivalent to rescaling the temperature as $T\leftarrow T (1 - \tilde g)$ since naturally $P_{\rm therm}\propto T$. 

As Fig.~\ref{fig:non-thermal-degen} shows in the case of a non-thermal contribution the $M_{\rm therm}$ measurement (which is carried out the same way as done by observations, i.e., assuming no non-thermal contribution) matches the lensing mass reasonably well \textit{in the case of modified gravity}. We achieved this by choosing the functional form of $\tilde g$ in Eq.~\eqref{eq:Pnontherm1} as
\begin{equation}
\tilde g(M) = \frac{1}{1 + \tilde a M_{13}^{\tilde\alpha}}\;,
\end{equation}
with $M_{13}\equiv M/(10^{13}M_\odot h^{-1})$ and $(\tilde a,\,\tilde \alpha)=(3/4,\,2/3)$.
This serves as an example of how unknown non-thermal physics can cancel out any signal originating from modified gravity -- which is a severe problem when trying to place constraints on the modifications of gravity using thermal measurements.

This problem will be alleviated once the contribution of non-thermal effects can be directly quantified using observational data (e.g., by measuring directly the intra-cluster turbulence). 
In the sequel of this subsection, we assume this has been done and is has been shown the contribution of the non-thermal components is negligible. We do this in order to show which constraints on modified gravity can be placed hypothetically using thermal mass estimates.

Fig.~\ref{fig:obs_constraints} and Fig.~\ref{fig:obs_constraints_fofr} show the resulting hypothetical constraints where we used the $M_{\rm thermal} / M_{\rm lens} \propto \gamma + 1$ relation described in \S~\ref{sec:comp-semi-analyt}. Note that in this case we did not consider a constant bias throughout the mass range which would shift the thermal mass measurements. Instead, we simply allowed for a maximum deviation from the (mean of the) measured mass ratio of $2\sigma$. For conversion to the $f(R)$ constraints we used -- as in the previous section -- our full simulation output (Fig.~\ref{fig:Mtherm}) and, thus, under the discussed assumptions found the rather conservative limit $|f_{R0}|<10^{-4}$ as presented in Fig.~\ref{fig:obs_constraints_fofr}.

\subsection{Caveats}
\label{sec:caveats}
Using clusters of galaxies to constrain modified gravity theories can be challenging as several sources of uncertainty have to be taken into account. From the observational side these uncertainties are immense for individual clusters but can be overcome when using a large number of objects -- if no effect alters the measured kinematic or thermal masses systematically. As this is uncertain in particular for the kinematic mass estimates we fixed the bias to a conservative value which should counter-act the effect\footnote{This procedure relies on the fact that these potential systematic effects (as well as other uncertainties) are captured by the observational error bars given.}. This leads, however, to the fact that near constant modifications of gravity throughout the whole measured mass range would not be detected. 

Another important cause of uncertainty is the theoretical modeling where -- although we included (basic) baryonic physics -- not all important physical effects are taken into account. For instance, it is expected that supernovea and AGN feedback mechanisms are somewhat degenerate with the enhancement of gravity and, thus, weaken constraints on modified gravity theories (see e.g. \cite{Puchwein2013MNRAS.436..348P,2016arXiv160202154M}).
This is in particular problematic for the thermal mass measurements as the `non-thermal pressure component' is not well understood theoretically as well as essentially completely unconstrained observationally. As explained in \S\ref{sec:therm-mass-meas} the non-thermal contribution is degenerate with the effect of modified gravity which makes the use of thermal measurements in order to constrain gravity only possible if there is independent measurements of the non-thermal contributions. Future missions will be able to measure the non-thermal pressure component and, hence, turn this systematic degeneracy into a factor with (potentially large) uncertainties \citep{2010A&A...510A..76L}.
Until then one has to resort to other probes (such as the kinematic mass) in order to constrain modified gravity theories. However, as these probes mainly rely on the dynamics of dark matter, models which are only coupled to baryons evade current constraints (see \S\ref{sec:results-non-universal}).

Overall, we want to stress that although clusters of galaxies are a powerful tool to constrain gravity on intermediate scales, also big uncertainties are associated with it which have to be dealt with. Nevertheless, they have the potential to close the gap between large-scale and local experiments as well as to probe the impact of gravity on dark-matter and baryons independently.

\section{Conclusions}
Using a hydrodynamic $N$-body code, we studied the effect of screened modified gravity models on the mass estimates of galaxy clusters. In particular, we focused on two novel aspects: \textit{(i)} we studied modified gravity models in which baryons and dark matter are coupled with different strengths to the scalar field, and, \textit{(ii)} we put the simulation results into the greater context of a general screened-modified gravity parametrization.

Our findings in these matters can be summarized as follows:
\begin{itemize}
\item The lensing mass of a cluster can differ tremendously from its kinematic or thermal mass in modified gravity theories. In screened modified gravity theories the magnitude of variation varies from a maximum to zero from the unscreened mass range to the screened one, respectively. This makes the mass measurements of clusters a powerful probe of gravity in different length scales and environments.
\item Differently coupled dark matter and baryons are hard to detect observationally as degeneracies exist. However, as the thermal mass is stronger affected by the baryonic coupling than the kinetic mass, possessing information about the three discussed mass estimates can break this degeneracy.
\item We placed the specific Symmetron and $f(R)$ models studied on a common parameter space which we also constrained using kinematic, lensing, and, thermal mass observations.
\item The ratio of the kinetic and lensing mass yields competitive constraints on the modification of gravity. Using thermal measurements, on the other hand, is currently unfeasible since the effect of non-thermal contributions is degenerate with a potential signal of modified gravity. This well be alleviated when these contributions are quantified in a model-independent way.
\end{itemize}
In conclusion, using various observational mass estimates for cluster of galaxies are a powerful tool in order to constrain modified gravity theories which possess a screening mechanism -- especially as future surveys increase the number of observed galaxies.

\begin{acknowledgements}
The authors thank the anonymous referee for the constructive comments that significantly improved the manuscript.
M.G. thanks the physics \& astronomy department at Johns Hopkins University for their kind hospitality. H.A.W. is supported by BIPAC and
the  Oxford  Martin  School. We thank the Research Council of Norway for their support. The simulations used in this paper were performed on the NOTUR cluster {\tt{HEXAGON}}, which is the computing facility at the University of Bergen.
\end{acknowledgements}

\appendix

\section{Including the non-thermal pressure component}
\label{sec:appendix}
It has been shown \citep{Evrard1990,Rasia2004,Kay2004,Dolag2005,Lau2009} that the pressure of the intracluster medium will have a significant non-thermal component generated by random gas motions and turbulence, so that the total pressure $P_{\rm Tot}$ of a cluster is
\begin{align}
P_{\rm Tot}(<r) = P_{\rm thermal}(r)+P_{\rm non-thermal}(r).
\end{align}
This results in the mass estimates will consist of a thermal and non-thermal component as well
\begin{align}
M(<r) = M_{\rm thermal}(r)+M_{\rm non-thermal}(r)
\end{align}
where 
\begin{align}
 M_{\rm thermal}(r) &= -\frac{r^2}{G\rho_{\rm gas}(r)}\frac{dP_{\rm thermal}(r)}{dr}\\
 M_{\rm non-thermal}(r) &= -\frac{r^2}{G\rho_{\rm gas}(r)}\frac{dP_{\rm non-thermal}(r)}{dr}.
\end{align}

\noindent By using $P_{\rm thermal}=kn_{\rm gas}T_{\rm gas}$, where $\rho_{\rm gas}=\mu m_p n_{\rm gas}$, we find that
\begin{align}
 \frac{dP_{\rm thermal}}{dr} = \frac{kT_{\rm gas}(r)}{\mu m_p}\left(\frac{d\rho_{\rm gas}(r)}{dr} + \frac{\rho_{\rm gas}(r)}{T_{\rm gas}(r)}\frac{dT_{\rm gas}(r)}{dr}\right)
\end{align}
so that
\begin{align}
M_{\rm therm} &= -\frac{k_Br^2T_{\rm thermal}(r)}{\mu m_p G}\left(\frac{d\ln\rho_{\rm thermal}}{dr} + \frac{d\ln T_{\rm thermal}}{dr}\right),
\end{align}
as show earlier in the paper.

Often, the non-thermal pressure is expressed as a fraction of the total pressure
\begin{align}
P_{\rm non-thermal}(r) = g(r)P_{\rm total}(r) = \frac{g(r)}{1-g(r)}P_{\rm thermal},
\end{align}
with the derivative 
\begin{align}
\frac{d P_{\rm non-thermal}}{dr} &= \frac{1}{1-g(r)}\left(g(r)\frac{dP_{\rm thermal}}{dr} + \frac{dg(r)}{dr}P_{\rm thermal} \right.\\
&\left.+ \frac{g(r)}{1-g(r)}\frac{dg(r)}{dr}P_{\rm thermal}\right).
\end{align} 

A fit to the $g$-function has been found in \citet{Shaw2010, Battaglia2012} who performed a series of 16 $\Lambda$CDM simulations to obtain
\begin{align}
 g(r) = \alpha_{\rm nt}(1+z)^{\beta_{\rm nt}}\left(\frac{r}{r_{500}}\right)^{n_{\rm nt}}\left(\frac{M_{200}}{3\times10^{14}M_{\odot}}\right)^{n_m},
\end{align}
where the free variables have the $\Lambda$CDM best-fit values $\alpha_{\rm nt}=0.18$, $\beta_{\rm nt}=0.5$, $n_{\rm nt}=0.8$, and $n_M=0.2$. The derivative of the $g$-factor is
\begin{align}
 \frac{dg(r)}{dr} = \frac{n_{\rm nt}}{r}g(r).
\end{align}

\begin{figure*}
  \includegraphics[width=.49\textwidth]{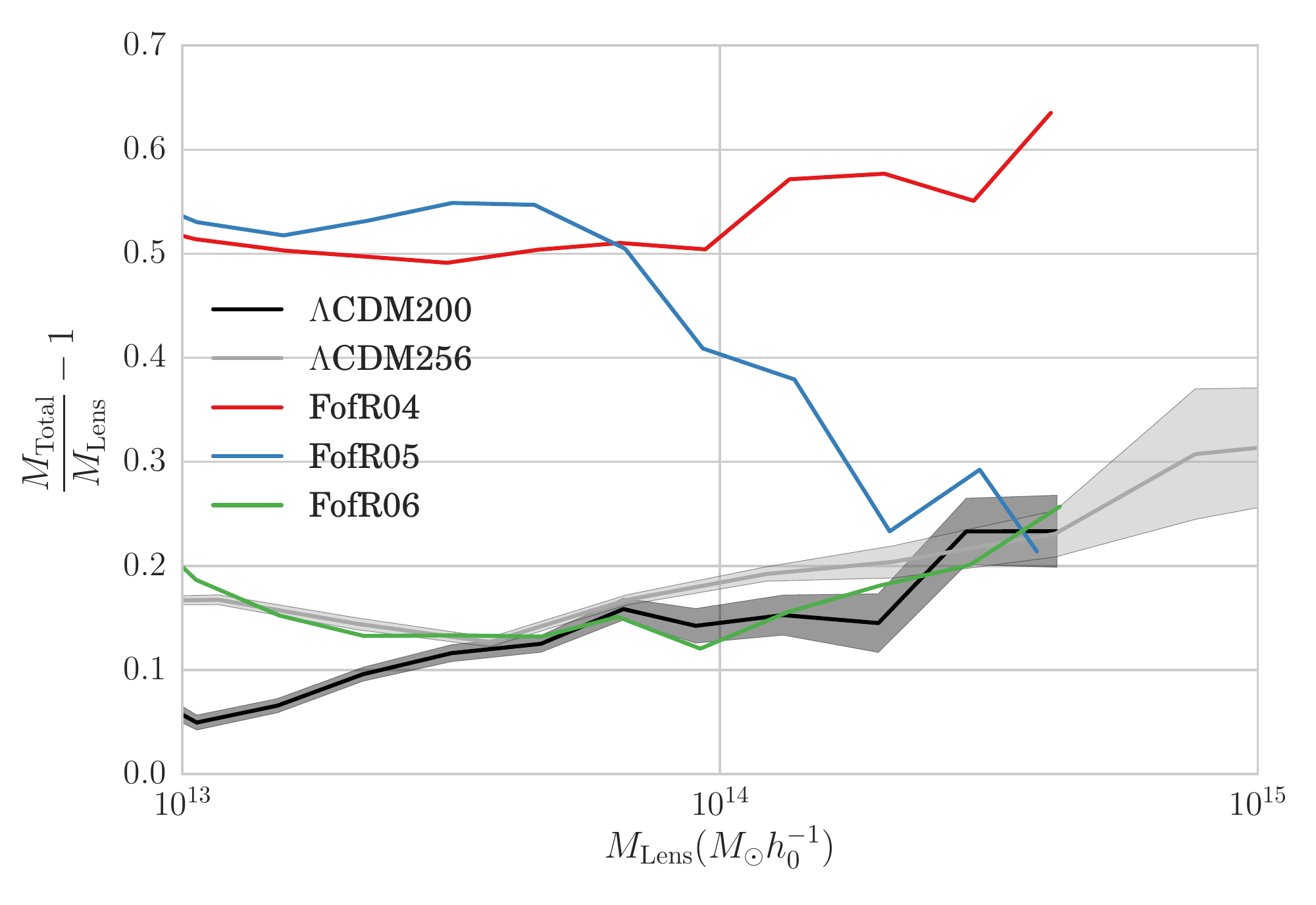}
  \includegraphics[width=.49\textwidth]{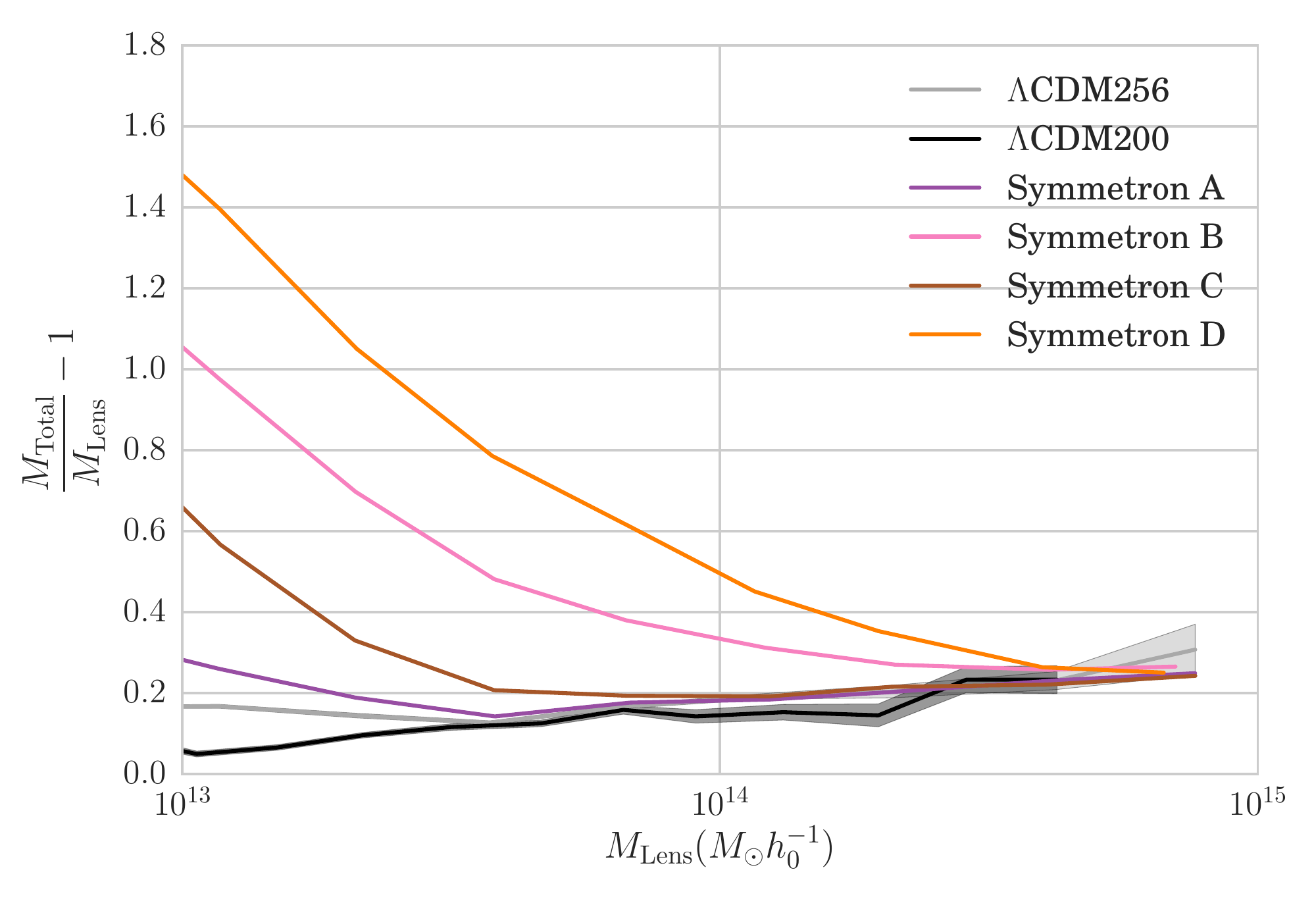}
  \caption{Ratio of the combined thermal and non-thermal mass and lensing mass for the analyzed $f(R)$ models (left panel) and Symmetron models (right panel). The horizontal lines and markers show the width and center of the transition region as defined in \S~\ref{sec:univ-param-scre}, respectively.}
  \label{fig:Nontherm}
\end{figure*}

Using the best fit we redo the analysis from before, now including the non-thermal pressure contribution, and present the results in \fig{fig:Nontherm}.

As we can see the results now differ substantially from \fig{fig:Mtherm}, with the non-thermal pressure component having introduced a strong mass dependence. However, we want to stress that this is just one particular example as the current expression of the non-thermal pressure contribution is derived from standard gravity simulations is strongly model dependent. Thus, we cannot simply use the expression as is for the modified gravity models. 

In spite of this complication, we want to note that \textit{in principle} it is possible to use the ratio between the thermal and lensing mass to constrain screened modified gravity theories, and also -- when including the kinetic mass -- to rule out certain combinations of non-universal coupling. All this, however, requires the contribution of the non-thermal pressure to be `under control', i.e., the magnitude of the intra-cluster turbulence are at least limited by observations.

\bibliography{references,references_fofr_const}

\begin{thebibliography}{102}
\expandafter\ifx\csname natexlab\endcsname\relax\def\natexlab#1{#1}\fi

\bibitem[{{Achitouv} {et~al.}(2015){Achitouv}, {Baldi}, {Puchwein}, \&
  {Weller}}]{2015arXiv151101494A}
{Achitouv}, I., {Baldi}, M., {Puchwein}, E., \& {Weller}, J. 2015, ArXiv
  e-prints [\eprint[arXiv]{1511.01494}]

\bibitem[{{Amendola} \& {Tsujikawa}(2010)}]{Amendola2010deto.book.....A}
{Amendola}, L. \& {Tsujikawa}, S. 2010, {Dark Energy: Theory and Observations}
  (Cambridge University Press)

\bibitem[{{Arnold} {et~al.}(2014){Arnold}, {Puchwein}, \&
  {Springel}}]{2014MNRAS.440..833A}
{Arnold}, C., {Puchwein}, E., \& {Springel}, V. 2014, \mnras, 440, 833

\bibitem[{{Babichev} {et~al.}(2009){Babichev}, {Deffayet}, \&
  {Ziour}}]{2009IJMPD..18.2147B}
{Babichev}, E., {Deffayet}, C., \& {Ziour}, R. 2009, International Journal of
  Modern Physics D, 18, 2147

\bibitem[{{Baker} {et~al.}(2015){Baker}, {Psaltis}, \&
  {Skordis}}]{Baker2015ApJ...802...63B}
{Baker}, T., {Psaltis}, D., \& {Skordis}, C. 2015, \apj, 802, 63

\bibitem[{{Barreira} {et~al.}(2015{\natexlab{a}}){Barreira}, {Bose}, \&
  {Li}}]{2015JCAP...12..059B}
{Barreira}, A., {Bose}, S., \& {Li}, B. 2015{\natexlab{a}}, \jcap, 12, 059

\bibitem[{{Barreira} {et~al.}(2013){Barreira}, {Li}, {Hellwing}, {Baugh}, \&
  {Pascoli}}]{2013JCAP...10..027B}
{Barreira}, A., {Li}, B., {Hellwing}, W.~A., {Baugh}, C.~M., \& {Pascoli}, S.
  2013, \jcap, 10, 027

\bibitem[{{Barreira} {et~al.}(2015{\natexlab{b}}){Barreira}, {Li}, {Jennings},
  {Merten}, {King}, {Baugh}, \& {Pascoli}}]{2015MNRAS.454.4085B}
{Barreira}, A., {Li}, B., {Jennings}, E., {et~al.} 2015{\natexlab{b}}, \mnras,
  454, 4085

\bibitem[{{Battaglia} {et~al.}(2012){Battaglia}, {Bond}, {Pfrommer}, \&
  {Sievers}}]{Battaglia2012}
{Battaglia}, N., {Bond}, J.~R., {Pfrommer}, C., \& {Sievers}, J.~L. 2012, \apj,
  758, 74

\bibitem[{{Becker} {et~al.}(2007){Becker}, {McKay}, {Koester}, {Wechsler},
  {Rozo}, {Evrard}, {Johnston}, {Sheldon}, {Annis}, {Lau}, {Nichol}, \&
  {Miller}}]{2007ApJ...669..905B}
{Becker}, M.~R., {McKay}, T.~A., {Koester}, B., {et~al.} 2007, \apj, 669, 905

\bibitem[{{Bertotti} {et~al.}(2003){Bertotti}, {Iess}, \&
  {Tortora}}]{Berotti2003Natur.425..374B}
{Bertotti}, B., {Iess}, L., \& {Tortora}, P. 2003, \nat, 425, 374

\bibitem[{{Bertschinger}(1999)}]{Grafic}
{Bertschinger}, E. 1999, {COSMICS: Cosmological initial conditions and
  microwave anisotropy codes}, astrophysics Source Code Library

\bibitem[{Bourliot {et~al.}(2007)Bourliot, Ferreira, Mota, \& Skordis}]{bour}
Bourliot, F., Ferreira, P.~G., Mota, D.~F., \& Skordis, C. 2007, Phys. Rev.,
  D75, 063508

\bibitem[{{Brax} {et~al.}(2012{\natexlab{a}}){Brax}, {Davis}, \&
  {Li}}]{2012PhLB..715...38B}
{Brax}, P., {Davis}, A.-C., \& {Li}, B. 2012{\natexlab{a}}, Physics Letters B,
  715, 38

\bibitem[{{Brax} {et~al.}(2012{\natexlab{b}}){Brax}, {Davis}, {Li}, \&
  {Winther}}]{2012PhRvD..86d4015B}
{Brax}, P., {Davis}, A.-C., {Li}, B., \& {Winther}, H.~A. 2012{\natexlab{b}},
  \prd, 86, 044015

\bibitem[{{Brax} \& {Valageas}(2014)}]{2014PhRvD..90b3507B}
{Brax}, P. \& {Valageas}, P. 2014, \prd, 90, 023507

\bibitem[{{Brax} {et~al.}(2004){Brax}, {van de Bruck}, {Davis}, {Khoury}, \&
  {Weltman}}]{2004PhRvD..70l3518B}
{Brax}, P., {van de Bruck}, C., {Davis}, A.-C., {Khoury}, J., \& {Weltman}, A.
  2004, \prd, 70, 123518

\bibitem[{{Bull} {et~al.}(2015){Bull}, {Akrami}, {Adamek}, {Baker}, {Bellini},
  {Beltr{\'a}n Jim{\'e}nez}, {Bentivegna}, {Camera}, {Clesse}, {Davis}, {Di
  Dio}, {Enander}, {Heavens}, {Heisenberg}, {Hu}, {Llinares}, {Maartens},
  {M{\"o}rtsell}, {Nadathur}, {Noller}, {Pasechnik}, {Pawlowski}, {Pereira},
  {Quartin}, {Ricciardone}, {Riemer-S{\o}rensen}, {Rinaldi}, {Sakstein},
  {Saltas}, {Salzano}, {Sawicki}, {Solomon}, {Spolyar}, {Starkman}, {Steer},
  {Tereno}, {Verde}, {Villaescusa-Navarro}, {von Strauss}, \&
  {Winther}}]{2015arXiv151205356B}
{Bull}, P., {Akrami}, Y., {Adamek}, J., {et~al.} 2015, ArXiv e-prints
  [\eprint[arXiv]{1512.05356}]

\bibitem[{Burrage \& Khoury(2014)}]{BurragePhysRevD.90.024001}
Burrage, C. \& Khoury, J. 2014, Phys. Rev. D, 90, 024001

\bibitem[{{Cataneo} {et~al.}(2015){Cataneo}, {Rapetti}, {Schmidt}, {Mantz},
  {Allen}, {Applegate}, {Kelly}, {von der Linden}, \&
  {Morris}}]{Cataneo2015PhRvD..92d4009C}
{Cataneo}, M., {Rapetti}, D., {Schmidt}, F., {et~al.} 2015, \prd, 92, 044009

\bibitem[{Clifton {et~al.}(2012)Clifton, Ferreira, Padilla, \&
  Skordis}]{Clifton2012}
Clifton, T., Ferreira, P.~G., Padilla, A., \& Skordis, C. 2012, \physrep, 513,
  1

\bibitem[{Clifton {et~al.}(2005)Clifton, Mota, \& Barrow}]{mota2}
Clifton, T., Mota, D.~F., \& Barrow, J.~D. 2005, Mon. Not. Roy. Astron. Soc.,
  358, 601

\bibitem[{{Corbett Moran} {et~al.}(2014){Corbett Moran}, {Teyssier}, \&
  {Li}}]{2014arXiv1408.2856C}
{Corbett Moran}, C., {Teyssier}, R., \& {Li}, B. 2014, ArXiv e-prints
  [\eprint[arXiv]{1408.2856}]

\bibitem[{{Damour} \& {Polyakov}(1994)}]{Damour1994NuPhB.423..532D}
{Damour}, T. \& {Polyakov}, A.~M. 1994, Nuclear Physics B, 423, 532

\bibitem[{{Davis} {et~al.}(2012){Davis}, {Li}, {Mota}, \&
  {Winther}}]{Davis2012ApJ...748...61D}
{Davis}, A.-C., {Li}, B., {Mota}, D.~F., \& {Winther}, H.~A. 2012, \apj, 748,
  61

\bibitem[{{de Rham}(2014)}]{2014LRR....17....7D}
{de Rham}, C. 2014, Living Reviews in Relativity, 17, 7

\bibitem[{{Dolag} {et~al.}(2005){Dolag}, {Vazza}, {Brunetti}, \&
  {Tormen}}]{Dolag2005}
{Dolag}, K., {Vazza}, F., {Brunetti}, G., \& {Tormen}, G. 2005, \mnras, 364,
  753

\bibitem[{{Dvali} {et~al.}(2000){Dvali}, {Gabadadze}, \&
  {Porrati}}]{2000PhLB..485..208D}
{Dvali}, G., {Gabadadze}, G., \& {Porrati}, M. 2000, Physics Letters B, 485,
  208

\bibitem[{{Evrard}(1990)}]{Evrard1990}
{Evrard}, A.~E. 1990, \apj, 363, 349

\bibitem[{{Evrard} {et~al.}(2008){Evrard}, {Bialek}, {Busha}, {White}, {Habib},
  {Heitmann}, {Warren}, {Rasia}, {Tormen}, {Moscardini}, {Power}, {Jenkins},
  {Gao}, {Frenk}, {Springel}, {White}, \& {Diemand}}]{Evrard2008}
{Evrard}, A.~E., {Bialek}, J., {Busha}, M., {et~al.} 2008, \apj, 672, 122

\bibitem[{{Fagernes Ivarsen} {et~al.}(2016){Fagernes Ivarsen}, {Bull},
  {Llinares}, \& {Mota}}]{2016arXiv160303072F}
{Fagernes Ivarsen}, M., {Bull}, P., {Llinares}, C., \& {Mota}, D.~F. 2016,
  ArXiv e-prints [\eprint[arXiv]{1603.03072}]

\bibitem[{{Falck} {et~al.}(2015){Falck}, {Koyama}, \&
  {Zhao}}]{Falck2015JCAP...07..049F}
{Falck}, B., {Koyama}, K., \& {Zhao}, G.-B. 2015, \jcap, 7, 49

\bibitem[{{Ferraro} {et~al.}(2011){Ferraro}, {Schmidt}, \&
  {Hu}}]{2011PhRvD..83f3503F}
{Ferraro}, S., {Schmidt}, F., \& {Hu}, W. 2011, \prd, 83, 063503

\bibitem[{Gannouji {et~al.}(2010)Gannouji, Moraes, Mota, Polarski, Tsujikawa,
  \& Winther}]{gan}
Gannouji, R., Moraes, B., Mota, D.~F., {et~al.} 2010, Phys. Rev., D82, 124006

\bibitem[{{Gil-Mar{\'{\i}}n} {et~al.}(2011){Gil-Mar{\'{\i}}n}, {Schmidt}, {Hu},
  {Jimenez}, \& {Verde}}]{2011JCAP...11..019G}
{Gil-Mar{\'{\i}}n}, H., {Schmidt}, F., {Hu}, W., {Jimenez}, R., \& {Verde}, L.
  2011, \jcap, 11, 019

\bibitem[{{Gronke} {et~al.}(2015{\natexlab{a}}){Gronke}, {Llinares}, {Mota}, \&
  {Winther}}]{Gronke2014b_dl}
{Gronke}, M., {Llinares}, C., {Mota}, D.~F., \& {Winther}, H.~A.
  2015{\natexlab{a}}, \mnras, 449, 2837

\bibitem[{{Gronke} {et~al.}(2015{\natexlab{b}}){Gronke}, {Mota}, \&
  {Winther}}]{2015arXiv150507129G}
{Gronke}, M., {Mota}, D.~F., \& {Winther}, H.~A. 2015{\natexlab{b}}, ArXiv
  e-prints [\eprint[arXiv]{1505.07129}]

\bibitem[{{Gronke} {et~al.}(2014){Gronke}, {Llinares}, \&
  {Mota}}]{Gronke2014_dl}
{Gronke}, M.~B., {Llinares}, C., \& {Mota}, D.~F. 2014, \aap, 562, A9

\bibitem[{{Hammami} {et~al.}(2015){Hammami}, {Llinares}, {Mota}, \&
  {Winther}}]{Hammami2015a}
{Hammami}, A., {Llinares}, C., {Mota}, D.~F., \& {Winther}, H.~A. 2015, \mnras,
  449, 3635

\bibitem[{{Hammami} \& {Mota}(2015)}]{Hammami2015b}
{Hammami}, A. \& {Mota}, D.~F. 2015, \aap, 584, A57

\bibitem[{{Hellwing} {et~al.}(2014){Hellwing}, {Barreira}, {Frenk}, {Li}, \&
  {Cole}}]{2014PhRvL.112v1102H}
{Hellwing}, W.~A., {Barreira}, A., {Frenk}, C.~S., {Li}, B., \& {Cole}, S.
  2014, Physical Review Letters, 112, 221102

\bibitem[{{Higuchi} \& {Shirasaki}(2016)}]{2016arXiv160301325H}
{Higuchi}, Y. \& {Shirasaki}, M. 2016, ArXiv e-prints
  [\eprint[arXiv]{1603.01325}]

\bibitem[{Hinterbichler \& Khoury(2010)}]{Hinterbichler}
Hinterbichler, K. \& Khoury, J. 2010, \prl, 104, 231301

\bibitem[{Hinterbichler {et~al.}(2011)Hinterbichler, Khoury, Levy, \&
  Matas}]{Hinterbichlera}
Hinterbichler, K., Khoury, J., Levy, A., \& Matas, A. 2011, \prd

\bibitem[{{Hu} {et~al.}(2013){Hu}, {Liguori}, {Bartolo}, \&
  {Matarrese}}]{2013PhRvD..88b4012H}
{Hu}, B., {Liguori}, M., {Bartolo}, N., \& {Matarrese}, S. 2013, \prd, 88,
  024012

\bibitem[{Hu \& Sawicki(2007)}]{Hu2007}
Hu, W. \& Sawicki, I. 2007, \prd, 1

\bibitem[{{Jain} \& {VanderPlas}(2011)}]{jain:11}
{Jain}, B. \& {VanderPlas}, J. 2011, \jcap, 10, 032

\bibitem[{{Jain} {et~al.}(2013){Jain}, {Vikram}, \& {Sakstein}}]{jain:13}
{Jain}, B., {Vikram}, V., \& {Sakstein}, J. 2013, \apj, 779, 39

\bibitem[{{Johnston} {et~al.}(2007){Johnston}, {Sheldon}, {Wechsler}, {Rozo},
  {Koester}, {Frieman}, {McKay}, {Evrard}, {Becker}, \&
  {Annis}}]{arXiv0709.1159J}
{Johnston}, D.~E., {Sheldon}, E.~S., {Wechsler}, R.~H., {et~al.} 2007, arXiv
  e-prints [\eprint[arXiv]{0709.1159}]

\bibitem[{Joyce {et~al.}(2014)Joyce, Jain, Khoury, \& Trodden}]{Joyce2014}
Joyce, A., Jain, B., Khoury, J., \& Trodden, M. 2014, arXiv e-prints
  [\eprint{arXiv:1407.0059}]

\bibitem[{{Kay} {et~al.}(2004){Kay}, {Thomas}, {Jenkins}, \&
  {Pearce}}]{Kay2004}
{Kay}, S.~T., {Thomas}, P.~A., {Jenkins}, A., \& {Pearce}, F.~R. 2004, \mnras,
  355, 1091

\bibitem[{Khoury(2010)}]{Khoury2010}
Khoury, J. 2010, arXiv e-prints [\eprint{arXiv:1011.5909}]

\bibitem[{{Khoury}(2013)}]{2013arXiv1312.2006K}
{Khoury}, J. 2013, arXiv e-prints [\eprint{arXiv:1312.2006}]

\bibitem[{{Khoury} \& {Weltman}(2004)}]{2004PhRvD..69d4026K}
{Khoury}, J. \& {Weltman}, A. 2004, \prd, 69, 044026

\bibitem[{Khoury \& Weltman(2004)}]{Khourya}
Khoury, J. \& Weltman, A. 2004, Phys. Rev. Lett., 93, 171104

\bibitem[{{Knollmann} \& {Knebe}(2009)}]{AHF}
{Knollmann}, S.~R. \& {Knebe}, A. 2009, \apjs, 182, 608

\bibitem[{{Koennig} {et~al.}(2014){Koennig}, {Akrami}, {Amendola}, {Motta}, \&
  {Solomon}}]{2014PhRvD..90l4014K}
{Koennig}, F., {Akrami}, Y., {Amendola}, L., {Motta}, M., \& {Solomon}, A.~R.
  2014, \prd, 90, 124014

\bibitem[{{Koyama}(2015)}]{2015arXiv150404623K}
{Koyama}, K. 2015, arXiv e-prints [\eprint[arXiv]{1504.04623}]

\bibitem[{{Lagan{\'a}} {et~al.}(2010){Lagan{\'a}}, {de Souza}, \&
  {Keller}}]{2010A&A...510A..76L}
{Lagan{\'a}}, T.~F., {de Souza}, R.~S., \& {Keller}, G.~R. 2010, \aap, 510, A76

\bibitem[{{Lam} {et~al.}(2012){Lam}, {Nishimichi}, {Schmidt}, \&
  {Takada}}]{2012PhRvL.109e1301L}
{Lam}, T.~Y., {Nishimichi}, T., {Schmidt}, F., \& {Takada}, M. 2012, Physical
  Review Letters, 109, 051301

\bibitem[{{Lau} {et~al.}(2009){Lau}, {Kravtsov}, \& {Nagai}}]{Lau2009}
{Lau}, E.~T., {Kravtsov}, A.~V., \& {Nagai}, D. 2009, \apj, 705, 1129

\bibitem[{Li {et~al.}(2011)Li, Mota, \& Barrow}]{mota4}
Li, B., Mota, D.~F., \& Barrow, J.~D. 2011, Astrophys. J., 728, 109

\bibitem[{{Li} {et~al.}(2013){Li}, {Zhao}, \& {Koyama}}]{2013JCAP...05..023L}
{Li}, B., {Zhao}, G.-B., \& {Koyama}, K. 2013, \jcap, 5, 23

\bibitem[{{Li} {et~al.}(2012){Li}, {Zhao}, {Teyssier}, \&
  {Koyama}}]{Li2012JCAP...01..051L}
{Li}, B., {Zhao}, G.-B., {Teyssier}, R., \& {Koyama}, K. 2012, \jcap, 1, 51

\bibitem[{{Llinares} \& {Mota}(2013)}]{2013PhRvL.110p1101L}
{Llinares}, C. \& {Mota}, D.~F. 2013, Physical Review Letters, 110, 161101

\bibitem[{Llinares {et~al.}(2014)Llinares, Mota, \& Winther}]{Llinares2014a}
Llinares, C., Mota, D.~F., \& Winther, H.~A. 2014, \aap, 562, A78

\bibitem[{{Lombriser}(2014)}]{2014AnP...526..259L}
{Lombriser}, L. 2014, Annalen der Physik, 526, 259

\bibitem[{{Lombriser} {et~al.}(2014){Lombriser}, {Koyama}, \&
  {Li}}]{lombriser:14}
{Lombriser}, L., {Koyama}, K., \& {Li}, B. 2014, \jcap, 3, 021

\bibitem[{{Lombriser} {et~al.}(2012){Lombriser}, {Schmidt}, {Baldauf},
  {Mandelbaum}, {Seljak}, \& {Smith}}]{2012PhRvD..85j2001L}
{Lombriser}, L., {Schmidt}, F., {Baldauf}, T., {et~al.} 2012, \prd, 85, 102001

\bibitem[{{Mahdavi} {et~al.}(2013){Mahdavi}, {Hoekstra}, {Babul}, {Bildfell},
  {Jeltema}, \& {Henry}}]{2013ApJ...767..116M}
{Mahdavi}, A., {Hoekstra}, H., {Babul}, A., {et~al.} 2013, \apj, 767, 116

\bibitem[{{Mead} {et~al.}(2016){Mead}, {Heymans}, {Lombriser}, {Peacock},
  {Steele}, \& {Winther}}]{2016arXiv160202154M}
{Mead}, A., {Heymans}, C., {Lombriser}, L., {et~al.} 2016, ArXiv e-prints
  [\eprint[arXiv]{1602.02154}]

\bibitem[{{Mead} {et~al.}(2015){Mead}, {Peacock}, {Lombriser}, \&
  {Li}}]{Mead2014}
{Mead}, A.~J., {Peacock}, J.~A., {Lombriser}, L., \& {Li}, B. 2015, \mnras,
  452, 4203

\bibitem[{{Mota} \& {Shaw}(2007)}]{2007PhRvD..75f3501M}
{Mota}, D.~F. \& {Shaw}, D.~J. 2007, \prd, 75, 063501

\bibitem[{{Munshi} {et~al.}(2014){Munshi}, {Hu}, {Renzi}, {Heavens}, \&
  {Coles}}]{2014MNRAS.442..821M}
{Munshi}, D., {Hu}, B., {Renzi}, A., {Heavens}, A., \& {Coles}, P. 2014,
  \mnras, 442, 821

\bibitem[{{Nicolis} {et~al.}(2009){Nicolis}, {Rattazzi}, \&
  {Trincherini}}]{2009PhRvD..79f4036N}
{Nicolis}, A., {Rattazzi}, R., \& {Trincherini}, E. 2009, \prd, 79, 064036

\bibitem[{{Oyaizu} {et~al.}(2008){Oyaizu}, {Lima}, \&
  {Hu}}]{2008PhRvD..78l3524O}
{Oyaizu}, H., {Lima}, M., \& {Hu}, W. 2008, \prd, 78, 123524

\bibitem[{{Perlmutter} {et~al.}(1999){Perlmutter}, {Aldering}, {Goldhaber},
  {Knop}, {Nugent}, {Castro}, {Deustua}, {Fabbro}, {Goobar}, {Groom}, {Hook},
  {Kim}, {Kim}, {Lee}, {Nunes}, {Pain}, {Pennypacker}, {Quimby}, {Lidman},
  {Ellis}, {Irwin}, {McMahon}, {Ruiz-Lapuente}, {Walton}, {Schaefer}, {Boyle},
  {Filippenko}, {Matheson}, {Fruchter}, {Panagia}, {Newberg}, {Couch}, \&
  {Project}}]{Perlmutter1999ApJ...517..565P}
{Perlmutter}, S., {Aldering}, G., {Goldhaber}, G., {et~al.} 1999, \apj, 517,
  565

\bibitem[{{Pfrommer}(2008)}]{2008MNRAS.385.1242P}
{Pfrommer}, C. 2008, \mnras, 385, 1242

\bibitem[{{Puchwein} {et~al.}(2013){Puchwein}, {Baldi}, \&
  {Springel}}]{Puchwein2013MNRAS.436..348P}
{Puchwein}, E., {Baldi}, M., \& {Springel}, V. 2013, \mnras, 436, 348

\bibitem[{{Rasia} {et~al.}(2004){Rasia}, {Tormen}, \& {Moscardini}}]{Rasia2004}
{Rasia}, E., {Tormen}, G., \& {Moscardini}, L. 2004, \mnras, 351, 237

\bibitem[{{Riess} {et~al.}(1998){Riess}, {Filippenko}, {Challis},
  {Clocchiatti}, {Diercks}, {Garnavich}, {Gilliland}, {Hogan}, {Jha},
  {Kirshner}, {Leibundgut}, {Phillips}, {Reiss}, {Schmidt}, {Schommer},
  {Smith}, {Spyromilio}, {Stubbs}, {Suntzeff}, \&
  {Tonry}}]{Riess1998AJ....116.1009R}
{Riess}, A.~G., {Filippenko}, A.~V., {Challis}, P., {et~al.} 1998, \aj, 116,
  1009

\bibitem[{{Schmidt}(2010)}]{2010PhRvD..81j3002S}
{Schmidt}, F. 2010, \prd, 81, 103002

\bibitem[{{Schmidt} {et~al.}(2009){Schmidt}, {Vikhlinin}, \&
  {Hu}}]{2009PhRvD..80h3505S}
{Schmidt}, F., {Vikhlinin}, A., \& {Hu}, W. 2009, \prd, 80, 083505

\bibitem[{{Shaw} {et~al.}(2010){Shaw}, {Nagai}, {Bhattacharya}, \&
  {Lau}}]{Shaw2010}
{Shaw}, L.~D., {Nagai}, D., {Bhattacharya}, S., \& {Lau}, E.~T. 2010, \apj,
  725, 1452

\bibitem[{{Sheldon} {et~al.}(2009){Sheldon}, {Johnston}, {Scranton}, {Koester},
  {McKay}, {Oyaizu}, {Cunha}, {Lima}, {Lin}, {Frieman}, {Wechsler}, {Annis},
  {Mandelbaum}, {Bahcall}, \& {Fukugita}}]{2009ApJ...703.2217S}
{Sheldon}, E.~S., {Johnston}, D.~E., {Scranton}, R., {et~al.} 2009, \apj, 703,
  2217

\bibitem[{{Smith}(2009)}]{Smith2009arXiv0907.4829S}
{Smith}, T.~L. 2009, ArXiv e-prints [\eprint[arXiv]{0907.4829}]

\bibitem[{{Terukina} {et~al.}(2014){Terukina}, {Lombriser}, {Yamamoto},
  {Bacon}, {Koyama}, \& {Nichol}}]{2014JCAP...04..013T}
{Terukina}, A., {Lombriser}, L., {Yamamoto}, K., {et~al.} 2014, \jcap, 4, 013

\bibitem[{{Tessore} {et~al.}(2015){Tessore}, {Winther}, {Metcalf}, {Ferreira},
  \& {Giocoli}}]{2015JCAP...10..036T}
{Tessore}, N., {Winther}, H.~A., {Metcalf}, R.~B., {Ferreira}, P.~G., \&
  {Giocoli}, C. 2015, \jcap, 10, 036

\bibitem[{{Teyssier}(2002)}]{2002A&A...385..337T}
{Teyssier}, R. 2002, \aap, 385, 337

\bibitem[{{Tinker} {et~al.}(2008){Tinker}, {Kravtsov}, {Klypin}, {Abazajian},
  {Warren}, {Yepes}, {Gottl{\"o}ber}, \& {Holz}}]{2008ApJ...688..709T}
{Tinker}, J., {Kravtsov}, A.~V., {Klypin}, A., {et~al.} 2008, \apj, 688, 709

\bibitem[{{Upadhye} \& {Steffen}(2013)}]{2013arXiv1306.6113U}
{Upadhye}, A. \& {Steffen}, J.~H. 2013, ArXiv e-prints
  [\eprint[arXiv]{1306.6113}]

\bibitem[{{Vainshtein}(1972)}]{1972PhLB...39..393V}
{Vainshtein}, A.~I. 1972, Physics Letters B, 39, 393

\bibitem[{{Vikram} {et~al.}(2013){Vikram}, {Cabr{\'e}}, {Jain}, \&
  {VanderPlas}}]{vikram:13}
{Vikram}, V., {Cabr{\'e}}, A., {Jain}, B., \& {VanderPlas}, J.~T. 2013, \jcap,
  8, 020

\bibitem[{{Wilcox} {et~al.}(2015){Wilcox}, {Bacon}, {Nichol}, {Rooney},
  {Terukina}, {Romer}, {Koyama}, {Zhao}, {Hood}, {Mann}, {Hilton},
  {Manolopoulou}, {Sahl{\'e}n}, {Collins}, {Liddle}, {Mayers}, {Mehrtens},
  {Miller}, {Stott}, \& {Viana}}]{2015MNRAS.452.1171W}
{Wilcox}, H., {Bacon}, D., {Nichol}, R.~C., {et~al.} 2015, \mnras, 452, 1171

\bibitem[{{Will}(2006)}]{Will2006LRR.....9....3W}
{Will}, C.~M. 2006, Living Reviews in Relativity, 9, 3

\bibitem[{{Williams} {et~al.}(2004){Williams}, {Turyshev}, \&
  {Boggs}}]{Williams2004PhRvL..93z1101W}
{Williams}, J.~G., {Turyshev}, S.~G., \& {Boggs}, D.~H. 2004, \prl, 93, 261101

\bibitem[{{Winther} \& {Ferreira}(2015)}]{2014arXiv1403.6492W}
{Winther}, H.~A. \& {Ferreira}, P.~G. 2015, \prd, 91, 123507

\bibitem[{{Winther} {et~al.}(2012){Winther}, {Mota}, \&
  {Li}}]{2012ApJ...756..166W}
{Winther}, H.~A., {Mota}, D.~F., \& {Li}, B. 2012, \apj, 756, 166

\bibitem[{{Winther} {et~al.}(2015){Winther}, {Schmidt}, {Barreira}, {Arnold},
  {Bose}, {Llinares}, {Baldi}, {Falck}, {Hellwing}, {Koyama}, {Li}, {Mota},
  {Puchwein}, {Smith}, \& {Zhao}}]{Winther2015arXiv150606384W}
{Winther}, H.~A., {Schmidt}, F., {Barreira}, A., {et~al.} 2015, ArXiv e-prints
  [\eprint[arXiv]{1506.06384}]

\bibitem[{{Zhang} {et~al.}(2010){Zhang}, {Okabe}, {Finoguenov}, {Smith},
  {Piffaretti}, {Valdarnini}, {Babul}, {Evrard}, {Mazzotta}, {Sanderson}, \&
  {Marrone}}]{2010ApJ...711.1033Z}
{Zhang}, Y.-Y., {Okabe}, N., {Finoguenov}, A., {et~al.} 2010, \apj, 711, 1033

\bibitem[{{Zu} {et~al.}(2014){Zu}, {Weinberg}, {Jennings}, {Li}, \&
  {Wyman}}]{2014MNRAS.445.1885Z}
{Zu}, Y., {Weinberg}, D.~H., {Jennings}, E., {Li}, B., \& {Wyman}, M. 2014,
  \mnras, 445, 1885

\bibitem[{Zumalacarregui {et~al.}(2010)Zumalacarregui, Koivisto, Mota, \&
  Ruiz-Lapuente}]{mota3}
Zumalacarregui, M., Koivisto, T.~S., Mota, D.~F., \& Ruiz-Lapuente, P. 2010,
  JCAP, 1005, 038

\end{thebibliography}

\end{document}